\documentclass[aps,prx,superscriptaddress,twocolumn,preprintnumbers,amsmath,amssymb]{revtex4-2}
\usepackage{color}
\usepackage{array}
\usepackage{amsmath}
\usepackage{amssymb}
\usepackage{graphicx}
\usepackage{pifont}
\usepackage{epstopdf}
\usepackage{bbm}
\usepackage{hyperref}
\usepackage{float}
\usepackage{bibentry}
\usepackage{tcolorbox}
\usepackage{bm}

\usepackage{braket}
\usepackage{simpler-wick}
\newcommand{\addJS}[1]{\textcolor{blue}{#1}}

\begin{document}
\title{Correlated quantum shift vector of particle-hole excitations} 
\author{Xu Yang}
\affiliation{Division of Physics and Applied Physics, School of Physical and Mathematical Sciences, Nanyang Technological University, Singapore 637371}
\author{Ajit Srivastava}
\affiliation{Department of Physics, Emory University, Atlanta, GA 30322, USA}
\affiliation{Department of Quantum Matter Physics, University of Geneva, Geneva 1211, Switzerland}
\author{Justin C. W. Song}
\email{justinsong@ntu.edu.sg}
\affiliation{Division of Physics and Applied Physics, School of Physical and Mathematical Sciences, Nanyang Technological University, Singapore 637371}

\begin{abstract}
Excitons are a prime example of how electron interactions affect optical response and excitation. We demonstrate that, beyond its spectra, the bound nature of an exciton's electron-hole pair produces a correlated quantum geometry: excitonic excitations possess a quantum shift vector that is independent of light polarization. We find this counterintuitive behavior has dramatic consequences for geometric response: e.g., in noncentrosymmetric but non-polar materials, vertical excitonic transitions possess vanishing shift vector zeroing their shift photocurrent; this contrasts with finite and strongly light polarization dependent shift vectors for non-interacting delocalized particle-hole excitations. This dichotomy makes shift vector a sharp diagnostic of the pair localization properties of particle-hole excitations and demonstrates the non-perturbative effects of
electron interactions in excited state quantum geometric response.
\end{abstract}
\maketitle

Particle-hole excitations are key processes in optoelectronics. 
Often characterized by their excitation spectra, the quantum geometry of (optical) transition dipole moments are now recognized to play an outsized role in dynamical response~\cite{ahn2022riemannian, ahn2020low_frequency, morimoto2016topological}: e.g., the quantum metric determines the spectral weight~\cite{ahn2022riemannian,onishi2024fundamental, onishi2025quantum} and the quantum shift vector~\cite{resta2024geometrical, morimoto2016topological}, that describes light-induced changes to electric polarization, produces bulk photocurrents even in the absence of p-n junctions~\cite{von1981theory, ahn2020low_frequency, ahn2022riemannian, morimoto2016topological, resta2024geometrical}.  
 
Weakly interacting particle-hole excitations posssess a quantum geometry that mirrors the wavefunction winding of single-particle Bloch states~\cite{ahn2020low_frequency, ahn2022riemannian}. This correspondence renders opto-electronic response powerful probes of the Bloch band geometry of quantum materials~\cite{ahn2022riemannian, ma2023photocurrent}. Excitonic particle-hole excitations, on the other hand, are fundamentally different. Even as each exciton is mobile and moves freely as a single unit, its electron and hole are bound strongly together by interactions producing a pair envelope wavefunction that is real-space localized~\cite{wannier1937structure}.

Here we argue that such electron-hole interactions drive a correlated excitonic quantum geometry distinct from free particle-hole excitations. In particular, we find the quantum shift vector for excitonic optical excitations is {\it independent} of the light polarization. This counterintuitive result sharply contrasts with free particle-hole excitations whose shift vectors are strongly dependent on light polarization and sign changing~\cite{wang2019ferroicity, xiong2021atomic}. 
It is especially striking given that both excitonic and delocalized particle-hole excitations can possess similar transition dipole moments and quantum weight. 

We trace this delineation to the interaction-induced relative coordinate localization of the electron-hole pair envelope wavefunction for bound excitons; free particle-hole pairs, in contrast, possess a delocalized pair envelope. We find pair localization differentiates the way geometric phases accumulate
enabling the quantum shift vector to act 
as a ``geometric ruler'' of pair localization. 

\begin{figure}
    \centering
    \includegraphics[width=1.0\linewidth]{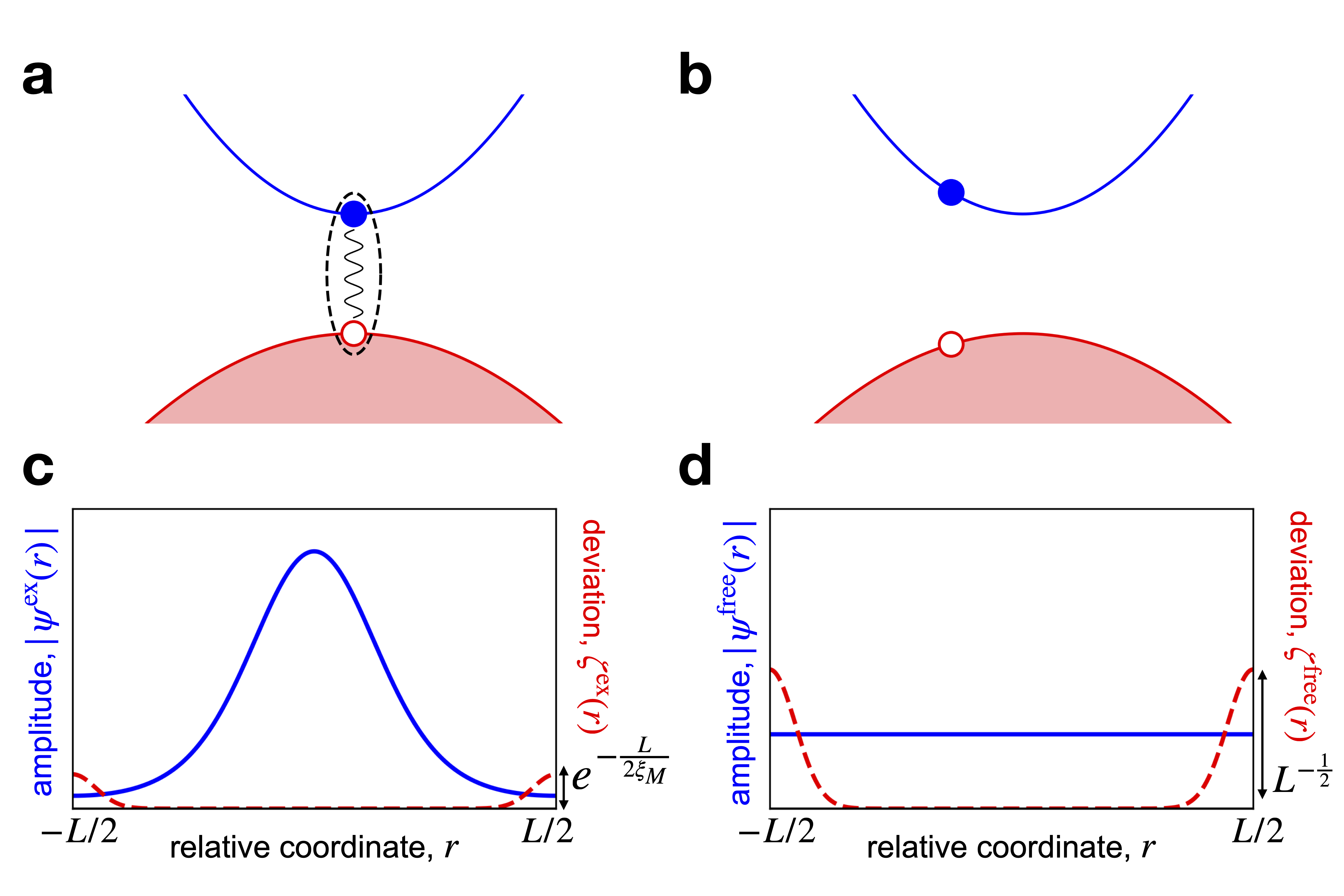}
    \caption{\textcolor{blue}{\it Localized vs. delocalized particle-hole excitations.} {\bf a.}  
  Both exciton and {\bf b.} free particle-hole excitations comprise electrons and holes in the conduction/valence bands. The former are localized by strong particle-hole interactions while the latter are delocalized. 
    {\bf c.} Schematic of amplitude of exciton envelope wavefunction 
    (blue) and flux-threading ansatz deviation $\zeta^{\text{ex}}(\textbf{r})$ (red) in an one-dimensional finite periodic system of size $L$. Because the exciton is bound, the deviation is exponentially suppressed at the boundaries. 
   {\bf d.} Schematic of amplitude of delocalized envelope of a free particle-hole pair (blue) and ``ansatz deviation'' $\zeta^{\text{free}}(\textbf{r})$ (red). For delocalized states, the deviation is large and comparable to that of the envelope function.}
    \label{fig:schematic}
\end{figure}

These correlated features readily dominate the photocurrent response for a range of technologically relevant optoelectronic materials~\cite{dong2023giant, akamatsu2021van}. A dramatic example are vertical transitions where we find the excitonic quantum shift vector vanishes in noncentrosymmetric but non-polar point groups rendering shift photocurrent zero for excitonic transitions. In sharp contrast, shift photocurrents for delocalized free particle-hole excitations are always allowed when inversion symmetry is broken, often display strong polarization dependence~\cite{xiong2021atomic, wang2019ferroicity}. This distinction naturally explains the dichotomy between recent experiments where zero bulk photocurrents for excitonic transitions were reported in noncentrosymmetric non-polar materials~\cite{akamatsu2021van, dong2023giant}
even while light polarization dependent and sign changing bulk photocurrrents are observed for free particle-hole excitations~\cite{krishna2025terahertz, ma2022intelligent}. Together, these underscore the importance of electron interactions in the optical response and quantum geometry of excited quantum matter.

\textcolor{blue}{{\it Flux threading and particle-hole excitations.}}
We begin by examining the structure of excitations of an insulating ground state, $\ket{{\rm GS}}$. For clarity, we focus on $\ket{{\rm GS}}$ with filled valence bands $v$ and empty conduction bands $c$.  
Generic particle-hole excitations on top of $\ket{{\rm GS}}$ with a conserved total momentum $\bm{Q}$ take the form: 
\begin{align}\label{eq:ph_basis}
    &\ket{\psi_{\bm{Q}}}_{\rm p-h}=\frac{1}{\sqrt{N}}\sum\limits_{\textbf{r},\textbf{R}_{\rm cm}}\psi_{\bm{Q}}(\textbf{r})e^{i\bm{Q}\cdot \textbf{R}_{\rm cm}}\ket{\textbf{R}_{\rm cm},\textbf{r}},
\end{align}
where $N$ is the number of unit-cells, $\ket{\textbf{R}_{\rm cm},\textbf{r}}\equiv c^{\dagger}_{c,\textbf{R}_{\rm cm}+\textbf{r}/2}c_{v,\textbf{R}_{\rm cm}-\textbf{r}/2}\ket{{\rm GS}}$ denotes a particle-hole pair with a center-of-mass coordinate $\textbf{R}_{\rm cm}$ and relative coordinate $\textbf{r}$. Here $c_{c(v),\textbf{R}}^\dag$ is the creation operator of a Wannier state with wavefunction $w_{c(v),\textbf{R}}(\textbf{x})$ \cite{martin2020electronic} in the conduction (valence) bands and $\psi_{\bm{Q}}(\textbf{r})$ is the envelope function of the particle-hole excitation. 
We will employ localized Wannier functions~\cite{marzari1997maximally,martin2020electronic}. 

Momentum $\bm{Q}$ is a good quantum number describing the free motion of the mobile center of mass of the particle-hole pair. In contrast, $\psi_{\bm{Q}}(\textbf{r})$ encodes the internal structure of the particle-hole excitation characterizing its interaction with electromagnetic fields. By projecting onto the particle-hole basis in Eq.~(\ref{eq:ph_basis}), $\psi_{\bm{Q}}(\textbf{r})$ can be directly obtained via the Bethe-Salpeter equation (BSE): $\sum\limits_{\textbf{r}'}\mathcal{H}_{\bm{Q}}(\textbf{r},\textbf{r}')\psi_{\bm{Q}}(\textbf{r}') = E\psi_{\bm{Q}}(\textbf{r})$ with \cite{wannier1937structure,mahan2013many,rohlfing2000electron}
\begin{equation}\label{eq:Bethe-Salpeter-eqn}
\mathcal{H}_{\bm{Q}} = [\epsilon_{c}(\hat{\textbf{p}}+\bm{Q}/2)-\epsilon_v(\hat{\textbf{p}}-\bm{Q}/2)] \delta_{\mathbf{r},\mathbf{r'}} + 
\mathcal{V}(\textbf{r},\textbf{r}'), 
\end{equation}
where $\hat{\textbf{p}}\equiv -i\partial_{\textbf{r}}$ and $\mathcal{V}(\textbf{r},\textbf{r}')$ is the particle-hole interaction; see {\bf Supplementary Information} for discussion. Here $\epsilon_{c,v}$ are the $c,v$ band energies.

In what follows, we focus on a real-space BSE to expose the critical role pair localization plays in the  correlated quantum geometry of excitons. Indeed, Eq.~(\ref{eq:Bethe-Salpeter-eqn}) readily displays how the scattering (free particle-hole excitations)  and bound (exciton) states are delineated by their pair localization properties in the {\it relative} (particle-hole) coordinate. When the excitation energies are above the optical gap, $\Delta (\bm{Q}) = \min_{\textbf{p}}[\epsilon_c (\mathbf{p} + \bm{Q}/2) - \epsilon_v(\mathbf{p} - \bm{Q}/2)]$, Eq.~(\ref{eq:Bethe-Salpeter-eqn}) can be naturally solved with 
the asymptotic plane wave scattering solutions for the envelope function ($e^{i\mathbf{p}\cdot\mathbf{r}}$ for large $r$) with energies $\epsilon_c(\mathbf{p} + \bm{Q}/2) - \epsilon_v(\mathbf{p} - \bm{Q}/2)$. Such scattering states are completely delocalized corresponding to unbound particle-hole pairs. In contrast, when the energy $E$ lies below the optical gap $\Delta (\bm{Q})$, 
the value of $\mathbf{p}$ becomes imaginary, leading to an exponentially decaying envelope function: these are the bound exciton states. 

As we now argue, the pair localization properties of the particle-hole excitation directly affect their quantum geometry. To illustrate this, we examine the behavior of the envelope function under insertion of uniform flux $\bm \kappa$ (units inverse length) in a periodic system size $L$. Flux insertion has become a useful framework for tracking the geometry and response of electronic systems~\cite{niu1985quantized,resta1998quantum,souza2000polarization,watanabe2018insensitivity}, for e.g., Wannier functions in Bloch systems transform as $w_{\textbf{R}}^{\kappa}(\textbf{x})=e^{-i\bm{\kappa}\cdot(\textbf{x}-\textbf{R})}w_{\textbf{R}}(\textbf{x})$. Flux-inserted Wannier functions
characterize the quantum geometry of electronic systems~\cite{luttinger1951effect,kohn1964theory,souza2000polarization}. For our case of particle-hole {\it pair} excitations in Eq.~(\ref{eq:Bethe-Salpeter-eqn}), we find fluxes accumulate through the relative coordinate as (see also {\bf SI}) 
\begin{equation}
\label{eq:kappaBSE}
\mathcal{H}_{\bm{Q}}^{\bm{\kappa}}(\textbf{r},\textbf{r}')=e^{-i\bm{\kappa}\cdot(\textbf{r}-\textbf{r}')}\mathcal{H}_{\bm{Q}}(\textbf{r},\textbf{r}'),
\end{equation}
where $\mathcal{H}_{\bm{Q}}^{\bm{\kappa}}$ is the flux-inserted BSE. 
Note that flux accumulation through the relative coordinate arises from opposite signs of particle and hole charge rendering the envelope function key to understanding its response. In contrast, for superconductors, flux couples to the center of mass coordinate of Cooper pairs \cite{matsyshyn2024superconducting}.

Importantly, the bound and scattering solutions of Eq.~(\ref{eq:Bethe-Salpeter-eqn}) change differently under flux insertion. The exponential pair localization of the exciton envelope function $\psi_{\bm{Q}}(\textbf{r})$ 
severely constrains its transformation properties. 
To see this, we apply a gauge transformation to gauge away the phase accumulated by flux insertion in Eq.~(\ref{eq:kappaBSE})  
\cite{kohn1964theory,souza2000polarization} via the ansatz:  
$\tilde{\psi}_{\bm{Q}}^\kappa(\textbf{r})=e^{-i\bm{\kappa}\cdot \textbf{r}}{\psi}_{\bm{Q}}(\textbf{r})$. 

Imporantly, notice that $\tilde{\psi}_{\bm{Q}}^\kappa(\textbf{r})$ is not the exact solution of Eq.~(\ref{eq:kappaBSE}). While it closely tracks the actual flux-inserted envelope wavefunction [solution of Eq.~(\ref{eq:kappaBSE})] in the bulk, it deviates close to the system boundary due to the phase factor $e^{-i\bm{\kappa}\cdot \textbf{r}}$ [see Fig.~\ref{fig:schematic}c,d]. This deviation can be quantified by direct substitution 
into Eq.~(\ref{eq:kappaBSE})(see evaluation in {\bf SI}): 
\begin{equation}\label{eq:error_estimation}
    \zeta(\bm{r})= \Big| \sum_{\textbf{r}'} \mathcal{H}^{\bm{\kappa}}_{\bm{Q}}(\textbf{r}, \textbf{r}') \tilde{\psi}^{\bm{\kappa}}_{\bm{Q}}(\textbf{r}') - E \tilde{\psi}^{\bm{\kappa}}_{\bm{Q}}(\textbf{r})\Big| \leq \mathcal{C}e^{-\frac{L}{2\xi_{M}}},
\end{equation}
where we have substituted the exciton envelope wavefunction in the last inequality: this exponentially suppresses the deviation [see Fig.~\ref{fig:schematic}c]. Here $\mathcal{C}$ is a constant independent of $L$ and $\xi_M$ is $\xi_{M}\equiv\text{max}(\xi_W,\xi)$, with $\xi_W$ and $\xi$ is the extent of the exponentially decaying Wannier function and exciton envelope function respectively.

\begin{figure*}[htbp!]
    \centering
    \includegraphics[width=1\linewidth]{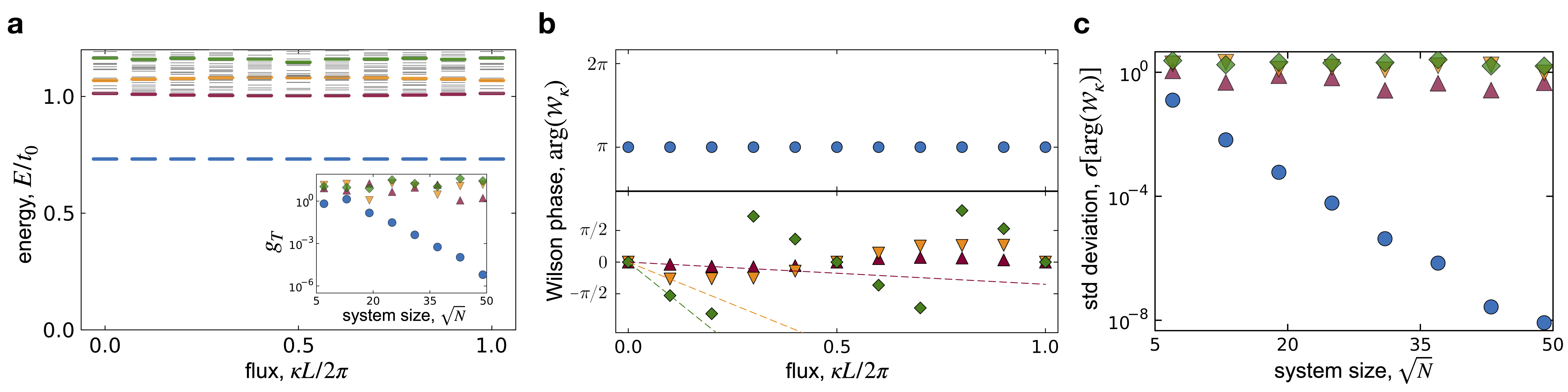}
    \caption{\textcolor{blue}{\it Numerical flux threading properties of particle-hole excitations.} 
    (a) Energy spectrum versus 
    flux $\kappa$:  exciton (blue) energy remains constant with flux, while delocalized particle-hole excitations vary significantly with $\kappa$ (gray, colored).  
    (inset) Thouless number $g_T \equiv |E^{\kappa L=0} - E^{\kappa L=\pi}| / (t_0/N)$ quantifies 
    sensitivity to flux insertion (normalized by  
    level spacing $t_0/N$): $g_T$ exponentially decays 
    for exciton (blue) but remains $\mathcal{O}(1)$ for free particle-hole pairs even at large $N$ (burgundy, yellow, green).  
    (b) Argument of Wilson loop operator (Eq.~\eqref{eq:shift_vector_difference}) with $\hat{V}_1 = (\sqrt{3}\hat{j}_x + \hat{j}_y)/2$ and $\hat{V}_2 = (\hat{j}_x - \sqrt{3}\hat{j}_y)/2$ versus  $\kappa$. For excitons, $\text{arg}(\mathcal{W}_{\bm \kappa})$ is flat 
    demonstrating insensitivity to $\hat{V}$;  
    for delocalized particle-hole excitations 
    $\text{arg}(\mathcal{W}_{\bm \kappa})$ shows strong $\bm{\kappa}$ dependence. Dashed lines indicate the derivative at $\bm{\kappa}=0$. (c) Standard deviation of $\text{arg}(\mathcal{W}_{\bm \kappa})$ vs.\ $\sqrt{N}$, showing exponential decay $\sim e^{-L/\xi_M}$ for exciton, and $\mathcal{O}(1)$ behavior for delocalized states. In all plots, Eq.~(\ref{eq:kappaBSE}) was numerically solved on an $N$ cell hexagonal lattice with staggered sublattice potential, system size $L=\sqrt{N}a$, and $a$ the lattice constant, see {\bf SI} for parameter values and numerical details.}
    \label{fig:numerical_result}
\end{figure*}

For excitons, $\zeta^{\rm ex} (\bm r)$ diminishes exponentially. 
Accordingly, we conclude that the bound excitonic state reads: 
\begin{equation}\label{eq:exciton_envelope_with_flux}
[\psi_{\bm{Q}}^{\bm{\kappa}}]^{\rm ex}(\textbf{r})=e^{-i\bm{\kappa}\cdot \textbf{r}}[{\psi}_{\bm{Q}}]^{\rm ex}(\textbf{r})+\mathcal{O}({\rm exp}\left[-L/(2\xi_{M})\right]), 
\end{equation}
with an energy that is insensitive to flux threading $E^{\rm ex}(\bm{\kappa})=E^{\rm ex}(\bm{0})+\mathcal{O}({\rm exp}\left[-L/\xi_{M}\right])$. Indeed, performing a scaling analysis in $L$ with numerical solutions of Eq.~(\ref{eq:kappaBSE}), we find the exciton's energy variation with $\bm \kappa$ is exponentially suppressed, Fig.~\ref{fig:numerical_result}{\bf a} and inset (blue). This mirrors that expected of an insulator~\cite{kohn1964theory}. 
As we will see below, this insensitivity to flux threading plays a key role in 
the polarization properties of excitonic excitations. 

In contrast, flux threading for delocalized free particle-hole excitations cannot be gauged away. While the deviation $\zeta^{\rm free} (\bm r)$ for the ansatz $e^{-i\bm{\kappa}\cdot\textbf{r}}{\psi}^{\rm free}_{\bm{Q}}(\textbf{r})$ vanishes in the bulk, it becomes significant near the boundary due to the discontinuous phase factor $e^{-i\bm{\kappa}\cdot\textbf{r}}$ across the boundary and the finite envelope amplitude there. The resulting violation is comparable in magnitude to the envelope function itself, see Fig.~\ref{fig:schematic}e. This indicates that
the delocalized states change in an essential way (not just via a phase factor) that is sensitive to the details of how $\bm{\kappa}$ is inserted. Indeed, a numerical solution of Eq.~(\ref{eq:kappaBSE}) for delocalized particle-hole excitations produces energy shifts comparable to the level spacing as $\bm \kappa$ is varied characteristic of a strong sensitivity to $\bm \kappa$, Fig.~\ref{fig:numerical_result}{\bf a}. 

\textcolor{blue}{{\it Shift vector probes \addJS{pair} localization of particle-hole excitations}.}
We now proceed to analyze the electric polarization change in a many-electron system produced by a particle-hole excitation. Such light induced changes are captured by a many-body quantum shift vector~\cite{resta2024geometrical}: 
\begin{equation}\label{eq:shift_vector}
\bm{\mathcal{R}}_{0\to n}\hspace{-1mm}\equiv\hspace{-1mm} 
\lim_{\bm \kappa \to 0} \big\{\delta\bm{\mathcal{A}}^{(0\to n)}_{\bm{\kappa}} 
\hspace{-1.5mm}+\hspace{-1mm} \nabla_{\bm{\kappa}}\text{arg}[\bra{\Phi_0 (\bm{\kappa})}\hat{V} (\bm{\kappa})\ket{\Phi_n (\bm{\kappa})}]\big\},
\end{equation} 
where $\ket{\Phi_{n,0} (\bm \kappa)}$ are the many-body excited and ground states,  
the many-body Berry connection difference reads 
$\delta \bm{\mathcal{A}}^{(0\to n)}_{\bm{\kappa}} = i\bra{\Phi_{n} (\bm{\kappa})}\nabla_{\bm{\kappa}}\ket{\Phi_{n} (\bm{\kappa})} - i\bra{\Phi_{0} (\bm{\kappa})}\nabla_{\bm{\kappa}}\ket{\Phi_{0} (\bm{\kappa})}$ for the excited $(n)$ and $(0)$ ground states, and $\hat{V}$ is the light-matter interaction. This flux-defined quantum shift vector~\cite{resta2024geometrical} allows to isolate the effect of real-space pair localization and address both free and excitonic excitations within the same framework.

Observe that all terms of Eq.~(\ref{eq:shift_vector}) are required to ensure its gauge invariance: the last term of Eq.~(\ref{eq:shift_vector}) compensates gauge transformations $\ket{\Phi} \to e^{i\theta(\bm{\kappa})} \ket{\Phi}$. As a result, light polarization dependence encoded in $\hat{V}$ is often regarded as an essential feature of light induced electric polarization changes~\cite{von1981theory,sipe2000second}; indeed, shift vectors often change sign with light polarization~\cite{xiong2021atomic}.  

We now show the converse is true for excitonic excitations:  polarization changes from an insulating many-body ground state to an excited excitonic state is {\it insensitive} to $\hat{V}$. To see this we consider two light matter interactions $\hat{V}_1$ and $\hat{V}_2$ e.g., from two different polarizations of light; both $\hat{V}_1$ and $\hat{V}_2$ produce an excitation from the ground to exciton state.  The difference between their light induced polarization changes $\delta \bm{\mathcal{R}} = \bm{\mathcal{R}}^{V_1}_{0\rightarrow {\rm ex}}-\bm{\mathcal{R}}^{V_2}_{0\rightarrow {\rm ex}}$ 
can be evaluated as: 
\begin{equation} \label{eq:shift_vector_difference}
    |\delta \bm{\mathcal{R}}| = |\nabla_{\bm{\kappa}}\text{arg} \mathcal{W}_{\bm{\kappa}}|
    < \mathcal{D} e^{-L/\xi_M}, 
\end{equation}
where $\mathcal{W}_{\bm{\kappa}} \hspace{-1mm}= \hspace{-1mm} \bra{\Phi_0 (\bm \kappa) }\hat{V}_1 (\bm \kappa)\ket{\Phi_{\rm ex} (\bm \kappa)}\bra{\Phi_{\rm ex} (\bm \kappa)}\hat{V}_2 (\bm \kappa)\ket{\Phi_0 (\bm \kappa)}$ is a Wilson loop~\cite{shi2019shift} tracking transition processes and $\mathcal{D}$ is a constant independent of $L$. 
Note in obtaining this bound we have applied Eq.~(\ref{eq:exciton_envelope_with_flux}) and the transformation properties of Wannier functions. Exponential suppression of $\delta \bm{\mathcal{R}}$ means that light induced changes to electric polarization from a ground state to an excitonic state is exponentially insensitive to $\hat{V}$; it loses all dependence on $\hat{V}$ in the thermodynamic limit $L\to \infty$. 

The behavior of the Wilson loop $\mathcal{W}_{\bm{\kappa}}$ as a function of $\bm \kappa$ for two different incident light polarizations is shown in Fig.~\ref{fig:numerical_result}(c). For excitonic states (upper panel), the $\bm \kappa$ dependence is exponentially suppressed, indicating that the shift vector is insensitive to light polarization. This stands in stark contrast to scattering states (lower panel), where $\mathcal{W}_{\bm{\kappa}}$ varies significantly with $\bm \kappa$: this reflects a strong light polarization dependence of the free particle-hole excitation shift vector~\cite{xiong2021atomic,wang2019ferroicity}. This sensitivity arises from the strong dependence of delocalized wavefunctions on flux insertion \cite{kohn1964theory,souza2000polarization}. As a result, we arrive at one of our key results: the shift vector probes the pair localization properties of particle-hole excitations.

\textcolor{blue}{{\it Symmetry and intrinsic excitonic transition shift vector.}} Note that while we have focused on light matter interaction $\hat{V}$, Eq.~(\ref{eq:shift_vector_difference}) applies generally for particle number conserving interactions, see {\bf SI}.  Indeed, by explicitly computing the polarization change between two excitonic states (e.g., transitions induced by disorder or phonons), we find: $\bm{\mathcal{R}}_{{\rm ex}_1\rightarrow {\rm ex}_2} \doteq \bm{\mathcal{R}}_{0\rightarrow {\rm ex}_2}-\bm{\mathcal{R}}_{0\rightarrow {\rm ex}_1}$ where ${\rm ex}_{1,2}$ denote distinct excitonic states (e.g., with different center of mass momentum $\bm{Q}$). Here we use $\doteq$ to denote equality up to an exponentially suppressed term in system size $L$ \cite{kohn1964theory}; it becomes an equality in the thermodynamic limit.

As a result, shift vectors for excitonic transitions are {\it intrinsic}: electric polarization changes induced by light only depend on the initial and final state. This produces a compact excitonic transition shift vector in real-space: 
\begin{align}\label{eq:shift_vector_intrinsic_expression}
\bm{\mathcal{R}}_{0\rightarrow {\rm ex}}\doteq\sum\limits_{\textbf{r},\textbf{r}'}\rho(\textbf{r},\textbf{r}')\big[\textbf{r}\delta_{\textbf{r},\textbf{r}'}+\bm{D}_{cv}(\textbf{r}'-\textbf{r}) \big],
\end{align}
where $\rho(\textbf{r},\textbf{r}')=[\psi^*_{\bm{Q}}]^{\rm ex}(\textbf{r})[\psi_{\bm{Q}}]^{\rm ex}(\textbf{r}')$ is the reduced density matrix in the relative coordinates and $\bm{D}_{cv}(\mathbf{R}) \equiv e^{i\bm{Q}\cdot\mathbf{R}/2}\bm{d}_c(\mathbf{R}) - e^{-i\bm{Q}\cdot\mathbf{R}/2}\bm{d}_v(\mathbf{R})$ represents the difference between the electron and hole dipole moments in the Wannier basis. Here $\bm{d}_{c(v)}(\mathbf{R}) \equiv \int d\mathbf{x} w^*_{c(v),0}(\mathbf{x}) \hat{\mathbf{x}} w_{c(v),\mathbf{R}}(\mathbf{x})$. A full discussion of the physical significance of each of these terms can be found in {\bf SI}. From a numerical perspective, Eq.~\eqref{eq:shift_vector_intrinsic_expression} enables efficient 
computation of the excitonic transition shift vector using real-space Wannier functions and exciton envelope functions from first-principles 
approaches~\cite{mostofi2008wannier90, haber2023maximally, davenport2024interaction, jankowski2025excitonic}.

We emphasize that $\bm{\mathcal{R}}_{0 \to \{ \cdots \} \to {\rm ex}}$ tracks the electric polarization change for excitonic transitions including the cumulative electronic effects of subsequent transitions between excitonic states. This is {\it a priori} distinct from the electric dipole of an exciton by itself~\cite{paiva2024shift}\footnote{$\bm{\mathcal{R}}_{0\to {\rm ex}}$ captures electric polarization changes produced by transitions (e.g., optically generated, scattering induced) closely tracking the properties of the relative coordinate important for shift responses. It should not be confused with the center-of-mass shift vector described in Ref.~\cite{paiva2024shift} that affects center-of-mass motion and dynamics.}. Strikingly, our analysis demonstrates that the details of the transitions are exponentially suppressed [Eq.~(\ref{eq:shift_vector_difference})] yielding $ \bm{\mathcal{R}}_{0 \to \{ \cdots\} \to {\rm ex}} \doteq \bm{\mathcal{R}}_{0 \to {\rm ex}}$. This counterintuitive result further delineates pair localized and delocalized excitations. 

The insensitivity of $\bm{\mathcal{R}}_{0\rightarrow {\rm ex}}$ to $\hat{V}$  
has important implications for its symmetry properties. Accounting for this, we find the excitonic transition shift vector transforms in the same way as $\nabla_{\bm{\kappa}}$ as a {\it vector} (see {\bf SI} for analysis):  
\begin{align}\label{eq:shift_vector_pg_transformation}
\bm{\mathcal{R}}_{0\rightarrow {\rm ex}} (\hat{U}_g\bm{Q})\doteq \hat{U}_g [\bm{\mathcal{R}}_{0\rightarrow {\rm ex}} (\bm Q)],
\end{align}
where $\hat{U}_g$ represents an operation of a point group symmetry (of the material) and we have explicitly specified center-of-mass $\bm{Q}$ for clarity. In contrast, $\bm{\mathcal{R}}_{0\rightarrow {\rm free}}$
does not transform as a vector~\cite{xiong2021atomic,wang2019ferroicity}. 

This distinction becomes dramatic for vertical transitions induced by far-field irradiation~\footnote{While finite momentum transfer $\bm{Q} \neq 0$ can enable to access non-vanishing finite $\bm{Q}$ shift vectors~\cite{shi2021geometric}, for far-field light $\bm{Q}$ are severely constrained by the light cone via energy-momentum conservation suppressing the finite $\bm Q$ shift vector}.
Taking $\bm Q =0$ in Eq.~(\ref{eq:shift_vector_pg_transformation}), we find that since $\bm{\mathcal{R}}_{0\rightarrow {\rm ex}}$ transforms as a vector it must be aligned along a polar axis. Importantly, it vanishes in non-polar point groups even when inversion symmetry is broken. In contrast, shift vector from delocalized particle-hole transitions remain finite in both polar and non-polar materials and point in a direction determined by both $\hat{V}$ 
and material properties~\cite{wang2019ferroicity, xiong2021atomic}. Notice that among 21 noncentrosymmetric point groups, only 10 are polar~\cite{aroyo2016international} yielding a point group dichotomy between shift vectors of excitonic and delocalized particle-hole excitations, see Table~\ref{tab:table1}.  

This feature directly impacts physical observables such as the many-body shift photocurrent~\cite{resta2024geometrical}: $\bm{j}_{\rm shift}=(-2\pi e/\hbar)\sum_n|\bra{n}\hat{V}^{\omega}\ket{0}|^2\bm{\mathcal{R}}_{0\to n}\delta(E_{n}-E_0-\hbar\omega)$ where $\hat{V}^\omega$ is the $\omega$ component of the light–matter interaction, see also {\bf SI}. For $\hbar \omega $ below the gap, it corresponds to excitonic transitions $n = {\rm ex}$. To appreciate the impact of Table~\ref{tab:table1}, consider the 
instructive example of 
two-dimensional semiconductors such as $C_{3z}$-symmetric 3R-MoS$_2$~\cite{dong2023giant}. Applying $C_{3z}$ point group operations onto Eq.~(\ref{eq:shift_vector_pg_transformation}) readily produces a {\it vanishing} in-plane ($x$–$y$) excitonic shift vector zeroing $\bm j_{\rm shift}$. In contrast, when $C_{3z}$ rotational symmetry is broken, a finite in-plane shift vector can manifest along the polar axis induced by strain allowing $\bm j_{\rm shift}$ to develop. This phenomenology directly matches the as-yet-unexplained experimental observations in Ref.~\cite{dong2023giant} where zero shift current was observed in $C_{3z}$-symmetric 3R-MoS$_2$; a giant shift current was activated when 3R-MoS$_2$ was strained. Our symmetry based analysis from Eq.~(\ref{eq:shift_vector_pg_transformation}) conclusively rules out shift photocurrents from excitonic transitions in non-polar systems thereby explaining this mystery~\cite{dong2023giant}.

\begin{table}[t]
\centering
\begin{tabular}{lcc}
\hline\hline
\textbf{PG Classes} & \textbf{    Excitonic shift       } & \textbf{Delocalized  shift  } \\
\hline
NCS, non-polar & \ding{55} & \ding{51} \\
NCS, polar     & \ding{51} & \ding{51} \\
\hline\hline
\end{tabular}
\caption{$\bm{Q}=0$ shift vector behavior for excitonic and delocalized states in different point group (PG) classes. NCS denotes noncentrosymmetric point groups.}
\label{tab:table1}
\end{table}

\textcolor{blue}{\it Exciton shift vector in topologically obstructed bands.} 
Our previous discussion relies on the exponentially decaying nature of Wannier functions. 
In topologically obstructed bands (e.g., $c,v$ bands with non-trivial Chern number), it is impossible to construct Wannier functions that are simultaneously exponentially localized along all spatial directions~\cite{brouder2007exponential}. 

This topological obstruction can be ``circumvented'', however, by employing hybrid Wannier functions, which are exponentially localized along one direction but delocalized along the others\cite{sgiarovello2001electron,marzari2012maximally}. Indeed, by aligning the pair localization direction with the flux $\bm{\kappa}$ (i.e. the direction of the shift), our conclusions remain valid namely: insensitivity of exciton envelope function to flux threading Eq.\eqref{eq:exciton_envelope_with_flux}, independence of exciton shift vector on light-matter interactions Eq.\eqref{eq:shift_vector_difference}, and the vector-like transformation rule of exciton shift vector under spatial symmetries Eq.\eqref{eq:shift_vector_pg_transformation}; see {\bf SI} for full discussion. 

\textcolor{blue}{\it Discussion.} Our work demonstrates that strong electron-hole interactions radically transform excited state quantum geometry: excitonic transition shift vectors are insensitive to light polarization and are intrinsic in nature. This result is non-perturbative and has important physical ramifications: excitonic transition shift current in non-polar materials vanishes, its light polarization dependence even in polar matter is suppressed, and shift vector is a ``geometric ruler'' for pair localization. 

From a computational perspective, the real space formulation of shift current responses we discuss only requires (i) the optical transition matrix element between the ground state and the {\it specific} excitonic state and 
(ii) the exciton shift vector in the real-space Wannier basis. Both are readily extracted from 
\emph{ab\;initio} GW+BSE calculations \cite{haber2023maximally}. This sharply contrasts with momentum-space 
formulations of the excitonic shift current~\cite{pedersen2015intraband,chan2021giant,ruan2023excitonic,morimoto2016exciton} that require an explicit summation over an entire set of intermediate states. We expect real-space calculations will open-up efficient numerics of shift current and other geometric responses~\cite{gao2020tunable, ma2023photocurrent} for excitons.

Perhaps most exciting is the prospect of electron interactions qualitatively transforming other aspects of quantum geometry~\cite{ahn2022riemannian, ahn2020low_frequency, morimoto2016topological, onishi2024fundamental, onishi2025quantum, resta2024geometrical} in strongly interacting materials. For e.g., the Hermitian curvature~\cite{ahn2022riemannian} of optical transitions involves covariant derivatives of the shift vector. Given the flux threading properties we introduce for excitonic transitions, we anticipate its Hermitian curvature may vanish 
producing a interaction-induced flat quantum geometry. Looking forward, we expect interactions will play critical roles in new classes of geometrical behavior beyond topology.

\textcolor{blue}{\it Acknowledgments.} We are grateful for helpful discussions with Goki Eda and Giovanni Vignale. This work was supported by the Singapore Ministry of Education (MOE) Academic Research Fund Tier 2 grant (MOE-T2EP50222-0011) and Tier 3 grant (MOE-MOET32023-0003) “Quantum Geometric Advantage”.

\textcolor{blue}{\it Author Contributions.} XY and JCWS conceived and planned the project. XY performed the calculations and numerical simulations with input from JCWS; AS contributed to discussions. XY and JCWS wrote the manuscript with input from AS. 

\bibliographystyle{apsrev4-2}
\bibliography{exciton}

\begin{thebibliography}{51}%
\makeatletter
\providecommand \@ifxundefined [1]{%
 \@ifx{#1\undefined}
}%
\providecommand \@ifnum [1]{%
 \ifnum #1\expandafter \@firstoftwo
 \else \expandafter \@secondoftwo
 \fi
}%
\providecommand \@ifx [1]{%
 \ifx #1\expandafter \@firstoftwo
 \else \expandafter \@secondoftwo
 \fi
}%
\providecommand \natexlab [1]{#1}%
\providecommand \enquote  [1]{``#1''}%
\providecommand \bibnamefont  [1]{#1}%
\providecommand \bibfnamefont [1]{#1}%
\providecommand \citenamefont [1]{#1}%
\providecommand \href@noop [0]{\@secondoftwo}%
\providecommand \href [0]{\begingroup \@sanitize@url \@href}%
\providecommand \@href[1]{\@@startlink{#1}\@@href}%
\providecommand \@@href[1]{\endgroup#1\@@endlink}%
\providecommand \@sanitize@url [0]{\catcode `\\12\catcode `\$12\catcode
  `\&12\catcode `\#12\catcode `\^12\catcode `\_12\catcode `\%12\relax}%
\providecommand \@@startlink[1]{}%
\providecommand \@@endlink[0]{}%
\providecommand \url  [0]{\begingroup\@sanitize@url \@url }%
\providecommand \@url [1]{\endgroup\@href {#1}{\urlprefix }}%
\providecommand \urlprefix  [0]{URL }%
\providecommand \Eprint [0]{\href }%
\providecommand \doibase [0]{https://doi.org/}%
\providecommand \selectlanguage [0]{\@gobble}%
\providecommand \bibinfo  [0]{\@secondoftwo}%
\providecommand \bibfield  [0]{\@secondoftwo}%
\providecommand \translation [1]{[#1]}%
\providecommand \BibitemOpen [0]{}%
\providecommand \bibitemStop [0]{}%
\providecommand \bibitemNoStop [0]{.\EOS\space}%
\providecommand \EOS [0]{\spacefactor3000\relax}%
\providecommand \BibitemShut  [1]{\csname bibitem#1\endcsname}%
\let\auto@bib@innerbib\@empty
\bibitem [{\citenamefont {Ahn}\ \emph {et~al.}(2022)\citenamefont {Ahn},
  \citenamefont {Guo}, \citenamefont {Nagaosa},\ and\ \citenamefont
  {Vishwanath}}]{ahn2022riemannian}%
  \BibitemOpen
  \bibfield  {author} {\bibinfo {author} {\bibfnamefont {J.}~\bibnamefont
  {Ahn}}, \bibinfo {author} {\bibfnamefont {G.-Y.}\ \bibnamefont {Guo}},
  \bibinfo {author} {\bibfnamefont {N.}~\bibnamefont {Nagaosa}},\ and\ \bibinfo
  {author} {\bibfnamefont {A.}~\bibnamefont {Vishwanath}},\ }\href@noop {}
  {\bibfield  {journal} {\bibinfo  {journal} {Nature Physics}\ }\textbf
  {\bibinfo {volume} {18}},\ \bibinfo {pages} {290} (\bibinfo {year}
  {2022})}\BibitemShut {NoStop}%
\bibitem [{\citenamefont {Ahn}\ \emph {et~al.}(2020)\citenamefont {Ahn},
  \citenamefont {Guo},\ and\ \citenamefont {Nagaosa}}]{ahn2020low_frequency}%
  \BibitemOpen
  \bibfield  {author} {\bibinfo {author} {\bibfnamefont {J.}~\bibnamefont
  {Ahn}}, \bibinfo {author} {\bibfnamefont {G.-Y.}\ \bibnamefont {Guo}},\ and\
  \bibinfo {author} {\bibfnamefont {N.}~\bibnamefont {Nagaosa}},\ }\href@noop
  {} {\bibfield  {journal} {\bibinfo  {journal} {Physical Review X}\ }\textbf
  {\bibinfo {volume} {10}},\ \bibinfo {pages} {041041} (\bibinfo {year}
  {2020})}\BibitemShut {NoStop}%
\bibitem [{\citenamefont {Morimoto}\ and\ \citenamefont
  {Nagaosa}(2016{\natexlab{a}})}]{morimoto2016topological}%
  \BibitemOpen
  \bibfield  {author} {\bibinfo {author} {\bibfnamefont {T.}~\bibnamefont
  {Morimoto}}\ and\ \bibinfo {author} {\bibfnamefont {N.}~\bibnamefont
  {Nagaosa}},\ }\href@noop {} {\bibfield  {journal} {\bibinfo  {journal}
  {Science advances}\ }\textbf {\bibinfo {volume} {2}},\ \bibinfo {pages}
  {e1501524} (\bibinfo {year} {2016}{\natexlab{a}})}\BibitemShut {NoStop}%
\bibitem [{\citenamefont {Onishi}\ and\ \citenamefont
  {Fu}(2024)}]{onishi2024fundamental}%
  \BibitemOpen
  \bibfield  {author} {\bibinfo {author} {\bibfnamefont {Y.}~\bibnamefont
  {Onishi}}\ and\ \bibinfo {author} {\bibfnamefont {L.}~\bibnamefont {Fu}},\
  }\href@noop {} {\bibfield  {journal} {\bibinfo  {journal} {Physical Review
  X}\ }\textbf {\bibinfo {volume} {14}},\ \bibinfo {pages} {011052} (\bibinfo
  {year} {2024})}\BibitemShut {NoStop}%
\bibitem [{\citenamefont {Onishi}\ and\ \citenamefont
  {Fu}(2025)}]{onishi2025quantum}%
  \BibitemOpen
  \bibfield  {author} {\bibinfo {author} {\bibfnamefont {Y.}~\bibnamefont
  {Onishi}}\ and\ \bibinfo {author} {\bibfnamefont {L.}~\bibnamefont {Fu}},\
  }\href@noop {} {\bibfield  {journal} {\bibinfo  {journal} {Physical Review
  Research}\ }\textbf {\bibinfo {volume} {7}},\ \bibinfo {pages} {023158}
  (\bibinfo {year} {2025})}\BibitemShut {NoStop}%
\bibitem [{\citenamefont {Resta}(2024)}]{resta2024geometrical}%
  \BibitemOpen
  \bibfield  {author} {\bibinfo {author} {\bibfnamefont {R.}~\bibnamefont
  {Resta}},\ }\href@noop {} {\bibfield  {journal} {\bibinfo  {journal}
  {Physical Review Letters}\ }\textbf {\bibinfo {volume} {133}},\ \bibinfo
  {pages} {206903} (\bibinfo {year} {2024})}\BibitemShut {NoStop}%
\bibitem [{\citenamefont {von Baltz}\ and\ \citenamefont
  {Kraut}(1981)}]{von1981theory}%
  \BibitemOpen
  \bibfield  {author} {\bibinfo {author} {\bibfnamefont {R.}~\bibnamefont {von
  Baltz}}\ and\ \bibinfo {author} {\bibfnamefont {W.}~\bibnamefont {Kraut}},\
  }\href@noop {} {\bibfield  {journal} {\bibinfo  {journal} {Physical Review
  B}\ }\textbf {\bibinfo {volume} {23}},\ \bibinfo {pages} {5590} (\bibinfo
  {year} {1981})}\BibitemShut {NoStop}%
\bibitem [{\citenamefont {Ma}\ \emph {et~al.}(2023)\citenamefont {Ma},
  \citenamefont {Krishna~Kumar}, \citenamefont {Xu}, \citenamefont {Koppens},\
  and\ \citenamefont {Song}}]{ma2023photocurrent}%
  \BibitemOpen
  \bibfield  {author} {\bibinfo {author} {\bibfnamefont {Q.}~\bibnamefont
  {Ma}}, \bibinfo {author} {\bibfnamefont {R.}~\bibnamefont {Krishna~Kumar}},
  \bibinfo {author} {\bibfnamefont {S.-Y.}\ \bibnamefont {Xu}}, \bibinfo
  {author} {\bibfnamefont {F.~H.}\ \bibnamefont {Koppens}},\ and\ \bibinfo
  {author} {\bibfnamefont {J.~C.}\ \bibnamefont {Song}},\ }\href@noop {}
  {\bibfield  {journal} {\bibinfo  {journal} {Nature Reviews Physics}\ }\textbf
  {\bibinfo {volume} {5}},\ \bibinfo {pages} {170} (\bibinfo {year}
  {2023})}\BibitemShut {NoStop}%
\bibitem [{\citenamefont {Wannier}(1937)}]{wannier1937structure}%
  \BibitemOpen
  \bibfield  {author} {\bibinfo {author} {\bibfnamefont {G.~H.}\ \bibnamefont
  {Wannier}},\ }\href@noop {} {\bibfield  {journal} {\bibinfo  {journal}
  {Physical Review}\ }\textbf {\bibinfo {volume} {52}},\ \bibinfo {pages} {191}
  (\bibinfo {year} {1937})}\BibitemShut {NoStop}%
\bibitem [{\citenamefont {Wang}\ and\ \citenamefont
  {Qian}(2019)}]{wang2019ferroicity}%
  \BibitemOpen
  \bibfield  {author} {\bibinfo {author} {\bibfnamefont {H.}~\bibnamefont
  {Wang}}\ and\ \bibinfo {author} {\bibfnamefont {X.}~\bibnamefont {Qian}},\
  }\href@noop {} {\bibfield  {journal} {\bibinfo  {journal} {Science advances}\
  }\textbf {\bibinfo {volume} {5}},\ \bibinfo {pages} {eaav9743} (\bibinfo
  {year} {2019})}\BibitemShut {NoStop}%
\bibitem [{\citenamefont {Xiong}\ \emph {et~al.}(2021)\citenamefont {Xiong},
  \citenamefont {Shi},\ and\ \citenamefont {Song}}]{xiong2021atomic}%
  \BibitemOpen
  \bibfield  {author} {\bibinfo {author} {\bibfnamefont {Y.}~\bibnamefont
  {Xiong}}, \bibinfo {author} {\bibfnamefont {L.-k.}\ \bibnamefont {Shi}},\
  and\ \bibinfo {author} {\bibfnamefont {J.~C.}\ \bibnamefont {Song}},\
  }\href@noop {} {\bibfield  {journal} {\bibinfo  {journal} {2D Materials}\
  }\textbf {\bibinfo {volume} {8}},\ \bibinfo {pages} {035008} (\bibinfo {year}
  {2021})}\BibitemShut {NoStop}%
\bibitem [{\citenamefont {Dong}\ \emph {et~al.}(2023)\citenamefont {Dong},
  \citenamefont {Yang}, \citenamefont {Yoshii}, \citenamefont {Matsuoka},
  \citenamefont {Kitamura}, \citenamefont {Hasegawa}, \citenamefont {Ogawa},
  \citenamefont {Morimoto}, \citenamefont {Ideue},\ and\ \citenamefont
  {Iwasa}}]{dong2023giant}%
  \BibitemOpen
  \bibfield  {author} {\bibinfo {author} {\bibfnamefont {Y.}~\bibnamefont
  {Dong}}, \bibinfo {author} {\bibfnamefont {M.-M.}\ \bibnamefont {Yang}},
  \bibinfo {author} {\bibfnamefont {M.}~\bibnamefont {Yoshii}}, \bibinfo
  {author} {\bibfnamefont {S.}~\bibnamefont {Matsuoka}}, \bibinfo {author}
  {\bibfnamefont {S.}~\bibnamefont {Kitamura}}, \bibinfo {author}
  {\bibfnamefont {T.}~\bibnamefont {Hasegawa}}, \bibinfo {author}
  {\bibfnamefont {N.}~\bibnamefont {Ogawa}}, \bibinfo {author} {\bibfnamefont
  {T.}~\bibnamefont {Morimoto}}, \bibinfo {author} {\bibfnamefont
  {T.}~\bibnamefont {Ideue}},\ and\ \bibinfo {author} {\bibfnamefont
  {Y.}~\bibnamefont {Iwasa}},\ }\href@noop {} {\bibfield  {journal} {\bibinfo
  {journal} {Nature nanotechnology}\ }\textbf {\bibinfo {volume} {18}},\
  \bibinfo {pages} {36} (\bibinfo {year} {2023})}\BibitemShut {NoStop}%
\bibitem [{\citenamefont {Akamatsu}\ \emph {et~al.}(2021)\citenamefont
  {Akamatsu}, \citenamefont {Ideue}, \citenamefont {Zhou}, \citenamefont
  {Dong}, \citenamefont {Kitamura}, \citenamefont {Yoshii}, \citenamefont
  {Yang}, \citenamefont {Onga}, \citenamefont {Nakagawa}, \citenamefont
  {Watanabe} \emph {et~al.}}]{akamatsu2021van}%
  \BibitemOpen
  \bibfield  {author} {\bibinfo {author} {\bibfnamefont {T.}~\bibnamefont
  {Akamatsu}}, \bibinfo {author} {\bibfnamefont {T.}~\bibnamefont {Ideue}},
  \bibinfo {author} {\bibfnamefont {L.}~\bibnamefont {Zhou}}, \bibinfo {author}
  {\bibfnamefont {Y.}~\bibnamefont {Dong}}, \bibinfo {author} {\bibfnamefont
  {S.}~\bibnamefont {Kitamura}}, \bibinfo {author} {\bibfnamefont
  {M.}~\bibnamefont {Yoshii}}, \bibinfo {author} {\bibfnamefont
  {D.}~\bibnamefont {Yang}}, \bibinfo {author} {\bibfnamefont {M.}~\bibnamefont
  {Onga}}, \bibinfo {author} {\bibfnamefont {Y.}~\bibnamefont {Nakagawa}},
  \bibinfo {author} {\bibfnamefont {K.}~\bibnamefont {Watanabe}}, \emph
  {et~al.},\ }\href@noop {} {\bibfield  {journal} {\bibinfo  {journal}
  {Science}\ }\textbf {\bibinfo {volume} {372}},\ \bibinfo {pages} {68}
  (\bibinfo {year} {2021})}\BibitemShut {NoStop}%
\bibitem [{\citenamefont {Krishna~Kumar}\ \emph {et~al.}(2025)\citenamefont
  {Krishna~Kumar}, \citenamefont {Li}, \citenamefont {Bertini}, \citenamefont
  {Chaudhary}, \citenamefont {Nowakowski}, \citenamefont {Park}, \citenamefont
  {Castilla}, \citenamefont {Zhan}, \citenamefont {Pantale{\'o}n},
  \citenamefont {Agarwal} \emph {et~al.}}]{krishna2025terahertz}%
  \BibitemOpen
  \bibfield  {author} {\bibinfo {author} {\bibfnamefont {R.}~\bibnamefont
  {Krishna~Kumar}}, \bibinfo {author} {\bibfnamefont {G.}~\bibnamefont {Li}},
  \bibinfo {author} {\bibfnamefont {R.}~\bibnamefont {Bertini}}, \bibinfo
  {author} {\bibfnamefont {S.}~\bibnamefont {Chaudhary}}, \bibinfo {author}
  {\bibfnamefont {K.}~\bibnamefont {Nowakowski}}, \bibinfo {author}
  {\bibfnamefont {J.~M.}\ \bibnamefont {Park}}, \bibinfo {author}
  {\bibfnamefont {S.}~\bibnamefont {Castilla}}, \bibinfo {author}
  {\bibfnamefont {Z.}~\bibnamefont {Zhan}}, \bibinfo {author} {\bibfnamefont
  {P.~A.}\ \bibnamefont {Pantale{\'o}n}}, \bibinfo {author} {\bibfnamefont
  {H.}~\bibnamefont {Agarwal}}, \emph {et~al.},\ }\href@noop {} {\bibfield
  {journal} {\bibinfo  {journal} {Nature Materials}\ ,\ \bibinfo {pages} {1}}
  (\bibinfo {year} {2025})}\BibitemShut {NoStop}%
\bibitem [{\citenamefont {Ma}\ \emph {et~al.}(2022)\citenamefont {Ma},
  \citenamefont {Yuan}, \citenamefont {Cheung}, \citenamefont {Watanabe},
  \citenamefont {Taniguchi}, \citenamefont {Zhang},\ and\ \citenamefont
  {Xia}}]{ma2022intelligent}%
  \BibitemOpen
  \bibfield  {author} {\bibinfo {author} {\bibfnamefont {C.}~\bibnamefont
  {Ma}}, \bibinfo {author} {\bibfnamefont {S.}~\bibnamefont {Yuan}}, \bibinfo
  {author} {\bibfnamefont {P.}~\bibnamefont {Cheung}}, \bibinfo {author}
  {\bibfnamefont {K.}~\bibnamefont {Watanabe}}, \bibinfo {author}
  {\bibfnamefont {T.}~\bibnamefont {Taniguchi}}, \bibinfo {author}
  {\bibfnamefont {F.}~\bibnamefont {Zhang}},\ and\ \bibinfo {author}
  {\bibfnamefont {F.}~\bibnamefont {Xia}},\ }\href@noop {} {\bibfield
  {journal} {\bibinfo  {journal} {Nature}\ }\textbf {\bibinfo {volume} {604}},\
  \bibinfo {pages} {266} (\bibinfo {year} {2022})}\BibitemShut {NoStop}%
\bibitem [{\citenamefont {Martin}(2020)}]{martin2020electronic}%
  \BibitemOpen
  \bibfield  {author} {\bibinfo {author} {\bibfnamefont {R.~M.}\ \bibnamefont
  {Martin}},\ }\href@noop {} {\emph {\bibinfo {title} {Electronic structure:
  basic theory and practical methods}}}\ (\bibinfo  {publisher} {Cambridge
  university press},\ \bibinfo {year} {2020})\BibitemShut {NoStop}%
\bibitem [{\citenamefont {Marzari}\ and\ \citenamefont
  {Vanderbilt}(1997)}]{marzari1997maximally}%
  \BibitemOpen
  \bibfield  {author} {\bibinfo {author} {\bibfnamefont {N.}~\bibnamefont
  {Marzari}}\ and\ \bibinfo {author} {\bibfnamefont {D.}~\bibnamefont
  {Vanderbilt}},\ }\href@noop {} {\bibfield  {journal} {\bibinfo  {journal}
  {Physical review B}\ }\textbf {\bibinfo {volume} {56}},\ \bibinfo {pages}
  {12847} (\bibinfo {year} {1997})}\BibitemShut {NoStop}%
\bibitem [{\citenamefont {Mahan}(2013)}]{mahan2013many}%
  \BibitemOpen
  \bibfield  {author} {\bibinfo {author} {\bibfnamefont {G.~D.}\ \bibnamefont
  {Mahan}},\ }\href@noop {} {\emph {\bibinfo {title} {Many-particle physics}}}\
  (\bibinfo  {publisher} {Springer Science \& Business Media},\ \bibinfo {year}
  {2013})\BibitemShut {NoStop}%
\bibitem [{\citenamefont {Rohlfing}\ and\ \citenamefont
  {Louie}(2000)}]{rohlfing2000electron}%
  \BibitemOpen
  \bibfield  {author} {\bibinfo {author} {\bibfnamefont {M.}~\bibnamefont
  {Rohlfing}}\ and\ \bibinfo {author} {\bibfnamefont {S.~G.}\ \bibnamefont
  {Louie}},\ }\href@noop {} {\bibfield  {journal} {\bibinfo  {journal}
  {Physical Review B}\ }\textbf {\bibinfo {volume} {62}},\ \bibinfo {pages}
  {4927} (\bibinfo {year} {2000})}\BibitemShut {NoStop}%
\bibitem [{\citenamefont {Niu}\ \emph {et~al.}(1985)\citenamefont {Niu},
  \citenamefont {Thouless},\ and\ \citenamefont {Wu}}]{niu1985quantized}%
  \BibitemOpen
  \bibfield  {author} {\bibinfo {author} {\bibfnamefont {Q.}~\bibnamefont
  {Niu}}, \bibinfo {author} {\bibfnamefont {D.~J.}\ \bibnamefont {Thouless}},\
  and\ \bibinfo {author} {\bibfnamefont {Y.-S.}\ \bibnamefont {Wu}},\
  }\href@noop {} {\bibfield  {journal} {\bibinfo  {journal} {Physical Review
  B}\ }\textbf {\bibinfo {volume} {31}},\ \bibinfo {pages} {3372} (\bibinfo
  {year} {1985})}\BibitemShut {NoStop}%
\bibitem [{\citenamefont {Resta}(1998)}]{resta1998quantum}%
  \BibitemOpen
  \bibfield  {author} {\bibinfo {author} {\bibfnamefont {R.}~\bibnamefont
  {Resta}},\ }\href@noop {} {\bibfield  {journal} {\bibinfo  {journal}
  {Physical Review Letters}\ }\textbf {\bibinfo {volume} {80}},\ \bibinfo
  {pages} {1800} (\bibinfo {year} {1998})}\BibitemShut {NoStop}%
\bibitem [{\citenamefont {Souza}\ \emph {et~al.}(2000)\citenamefont {Souza},
  \citenamefont {Wilkens},\ and\ \citenamefont
  {Martin}}]{souza2000polarization}%
  \BibitemOpen
  \bibfield  {author} {\bibinfo {author} {\bibfnamefont {I.}~\bibnamefont
  {Souza}}, \bibinfo {author} {\bibfnamefont {T.}~\bibnamefont {Wilkens}},\
  and\ \bibinfo {author} {\bibfnamefont {R.~M.}\ \bibnamefont {Martin}},\
  }\href@noop {} {\bibfield  {journal} {\bibinfo  {journal} {Physical Review
  B}\ }\textbf {\bibinfo {volume} {62}},\ \bibinfo {pages} {1666} (\bibinfo
  {year} {2000})}\BibitemShut {NoStop}%
\bibitem [{\citenamefont {Watanabe}(2018)}]{watanabe2018insensitivity}%
  \BibitemOpen
  \bibfield  {author} {\bibinfo {author} {\bibfnamefont {H.}~\bibnamefont
  {Watanabe}},\ }\href@noop {} {\bibfield  {journal} {\bibinfo  {journal}
  {Physical Review B}\ }\textbf {\bibinfo {volume} {98}},\ \bibinfo {pages}
  {155137} (\bibinfo {year} {2018})}\BibitemShut {NoStop}%
\bibitem [{\citenamefont {Luttinger}(1951)}]{luttinger1951effect}%
  \BibitemOpen
  \bibfield  {author} {\bibinfo {author} {\bibfnamefont {J.}~\bibnamefont
  {Luttinger}},\ }\href@noop {} {\bibfield  {journal} {\bibinfo  {journal}
  {Physical Review}\ }\textbf {\bibinfo {volume} {84}},\ \bibinfo {pages} {814}
  (\bibinfo {year} {1951})}\BibitemShut {NoStop}%
\bibitem [{\citenamefont {Kohn}(1964)}]{kohn1964theory}%
  \BibitemOpen
  \bibfield  {author} {\bibinfo {author} {\bibfnamefont {W.}~\bibnamefont
  {Kohn}},\ }\href@noop {} {\bibfield  {journal} {\bibinfo  {journal} {Physical
  review}\ }\textbf {\bibinfo {volume} {133}},\ \bibinfo {pages} {A171}
  (\bibinfo {year} {1964})}\BibitemShut {NoStop}%
\bibitem [{\citenamefont {Matsyshyn}\ \emph {et~al.}(2024)\citenamefont
  {Matsyshyn}, \citenamefont {Vignale},\ and\ \citenamefont
  {Song}}]{matsyshyn2024superconducting}%
  \BibitemOpen
  \bibfield  {author} {\bibinfo {author} {\bibfnamefont {O.}~\bibnamefont
  {Matsyshyn}}, \bibinfo {author} {\bibfnamefont {G.}~\bibnamefont {Vignale}},\
  and\ \bibinfo {author} {\bibfnamefont {J.~C.}\ \bibnamefont {Song}},\
  }\href@noop {} {\bibfield  {journal} {\bibinfo  {journal} {arXiv preprint
  arXiv:2410.21363}\ } (\bibinfo {year} {2024})}\BibitemShut {NoStop}%
\bibitem [{\citenamefont {Sipe}\ and\ \citenamefont
  {Shkrebtii}(2000)}]{sipe2000second}%
  \BibitemOpen
  \bibfield  {author} {\bibinfo {author} {\bibfnamefont {J.}~\bibnamefont
  {Sipe}}\ and\ \bibinfo {author} {\bibfnamefont {A.}~\bibnamefont
  {Shkrebtii}},\ }\href@noop {} {\bibfield  {journal} {\bibinfo  {journal}
  {Physical Review B}\ }\textbf {\bibinfo {volume} {61}},\ \bibinfo {pages}
  {5337} (\bibinfo {year} {2000})}\BibitemShut {NoStop}%
\bibitem [{\citenamefont {Shi}\ and\ \citenamefont
  {Song}(2019)}]{shi2019shift}%
  \BibitemOpen
  \bibfield  {author} {\bibinfo {author} {\bibfnamefont {L.-k.}\ \bibnamefont
  {Shi}}\ and\ \bibinfo {author} {\bibfnamefont {J.~C.}\ \bibnamefont {Song}},\
  }\href@noop {} {\bibfield  {journal} {\bibinfo  {journal} {Physical Review
  B}\ }\textbf {\bibinfo {volume} {100}},\ \bibinfo {pages} {201405} (\bibinfo
  {year} {2019})}\BibitemShut {NoStop}%
\bibitem [{\citenamefont {Mostofi}\ \emph {et~al.}(2008)\citenamefont
  {Mostofi}, \citenamefont {Yates}, \citenamefont {Lee}, \citenamefont {Souza},
  \citenamefont {Vanderbilt},\ and\ \citenamefont
  {Marzari}}]{mostofi2008wannier90}%
  \BibitemOpen
  \bibfield  {author} {\bibinfo {author} {\bibfnamefont {A.~A.}\ \bibnamefont
  {Mostofi}}, \bibinfo {author} {\bibfnamefont {J.~R.}\ \bibnamefont {Yates}},
  \bibinfo {author} {\bibfnamefont {Y.-S.}\ \bibnamefont {Lee}}, \bibinfo
  {author} {\bibfnamefont {I.}~\bibnamefont {Souza}}, \bibinfo {author}
  {\bibfnamefont {D.}~\bibnamefont {Vanderbilt}},\ and\ \bibinfo {author}
  {\bibfnamefont {N.}~\bibnamefont {Marzari}},\ }\href@noop {} {\bibfield
  {journal} {\bibinfo  {journal} {Computer physics communications}\ }\textbf
  {\bibinfo {volume} {178}},\ \bibinfo {pages} {685} (\bibinfo {year}
  {2008})}\BibitemShut {NoStop}%
\bibitem [{\citenamefont {Haber}\ \emph {et~al.}(2023)\citenamefont {Haber},
  \citenamefont {Qiu}, \citenamefont {da~Jornada},\ and\ \citenamefont
  {Neaton}}]{haber2023maximally}%
  \BibitemOpen
  \bibfield  {author} {\bibinfo {author} {\bibfnamefont {J.~B.}\ \bibnamefont
  {Haber}}, \bibinfo {author} {\bibfnamefont {D.~Y.}\ \bibnamefont {Qiu}},
  \bibinfo {author} {\bibfnamefont {F.~H.}\ \bibnamefont {da~Jornada}},\ and\
  \bibinfo {author} {\bibfnamefont {J.~B.}\ \bibnamefont {Neaton}},\
  }\href@noop {} {\bibfield  {journal} {\bibinfo  {journal} {Physical Review
  B}\ }\textbf {\bibinfo {volume} {108}},\ \bibinfo {pages} {125118} (\bibinfo
  {year} {2023})}\BibitemShut {NoStop}%
\bibitem [{\citenamefont {Davenport}\ \emph {et~al.}(2024)\citenamefont
  {Davenport}, \citenamefont {Knolle},\ and\ \citenamefont
  {Schindler}}]{davenport2024interaction}%
  \BibitemOpen
  \bibfield  {author} {\bibinfo {author} {\bibfnamefont {H.}~\bibnamefont
  {Davenport}}, \bibinfo {author} {\bibfnamefont {J.}~\bibnamefont {Knolle}},\
  and\ \bibinfo {author} {\bibfnamefont {F.}~\bibnamefont {Schindler}},\
  }\href@noop {} {\bibfield  {journal} {\bibinfo  {journal} {Physical Review
  Letters}\ }\textbf {\bibinfo {volume} {133}},\ \bibinfo {pages} {176601}
  (\bibinfo {year} {2024})}\BibitemShut {NoStop}%
\bibitem [{\citenamefont {Jankowski}\ \emph {et~al.}(2025)\citenamefont
  {Jankowski}, \citenamefont {Thompson}, \citenamefont {Monserrat},\ and\
  \citenamefont {Slager}}]{jankowski2025excitonic}%
  \BibitemOpen
  \bibfield  {author} {\bibinfo {author} {\bibfnamefont {W.~J.}\ \bibnamefont
  {Jankowski}}, \bibinfo {author} {\bibfnamefont {J.~J.}\ \bibnamefont
  {Thompson}}, \bibinfo {author} {\bibfnamefont {B.}~\bibnamefont
  {Monserrat}},\ and\ \bibinfo {author} {\bibfnamefont {R.-J.}\ \bibnamefont
  {Slager}},\ }\href@noop {} {\bibfield  {journal} {\bibinfo  {journal} {Nature
  Communications}\ }\textbf {\bibinfo {volume} {16}},\ \bibinfo {pages} {1}
  (\bibinfo {year} {2025})}\BibitemShut {NoStop}%
\bibitem [{\citenamefont {Paiva}\ \emph {et~al.}(2024)\citenamefont {Paiva},
  \citenamefont {Holder},\ and\ \citenamefont {Ilan}}]{paiva2024shift}%
  \BibitemOpen
  \bibfield  {author} {\bibinfo {author} {\bibfnamefont {C.}~\bibnamefont
  {Paiva}}, \bibinfo {author} {\bibfnamefont {T.}~\bibnamefont {Holder}},\ and\
  \bibinfo {author} {\bibfnamefont {R.}~\bibnamefont {Ilan}},\ }\href@noop {}
  {\bibfield  {journal} {\bibinfo  {journal} {arXiv preprint arXiv:2408.10300}\
  } (\bibinfo {year} {2024})}\BibitemShut {NoStop}%
\bibitem [{Note1()}]{Note1}%
  \BibitemOpen
  \bibinfo {note} {$\protect \bm {\protect \mathcal {R}}_{0\to {\protect \rm
  ex}}$ captures electric polarization changes produced by transitions (e.g.,
  optically generated, scattering induced) closely tracking the properties of
  the relative coordinate important for shift responses. It should not be
  confused with the center-of-mass shift vector described in Ref.~\cite
  {paiva2024shift} that affects center-of-mass motion and
  dynamics.}\BibitemShut {Stop}%
\bibitem [{Note2()}]{Note2}%
  \BibitemOpen
  \bibinfo {note} {While finite momentum transfer $\protect \bm {Q} \protect
  \neq 0$ can enable to access non-vanishing finite $\protect \bm {Q}$ shift
  vectors~\cite {shi2021geometric}, for far-field light $\protect \bm {Q}$ are
  severely constrained by the light cone via energy-momentum conservation
  suppressing the finite $\protect \bm {Q}$ shift vector}\BibitemShut {NoStop}%
\bibitem [{\citenamefont {Aroyo}(2016)}]{aroyo2016international}%
  \BibitemOpen
  \bibinfo {editor} {\bibfnamefont {M.~I.}\ \bibnamefont {Aroyo}},\ ed.,\
  \href@noop {} {\emph {\bibinfo {title} {International Tables for
  Crystallography, Volume A: Space-Group Symmetry}}}\ (\bibinfo  {publisher}
  {Wiley Online Library},\ \bibinfo {year} {2016})\BibitemShut {NoStop}%
\bibitem [{\citenamefont {Brouder}\ \emph {et~al.}(2007)\citenamefont
  {Brouder}, \citenamefont {Panati}, \citenamefont {Calandra}, \citenamefont
  {Mourougane},\ and\ \citenamefont {Marzari}}]{brouder2007exponential}%
  \BibitemOpen
  \bibfield  {author} {\bibinfo {author} {\bibfnamefont {C.}~\bibnamefont
  {Brouder}}, \bibinfo {author} {\bibfnamefont {G.}~\bibnamefont {Panati}},
  \bibinfo {author} {\bibfnamefont {M.}~\bibnamefont {Calandra}}, \bibinfo
  {author} {\bibfnamefont {C.}~\bibnamefont {Mourougane}},\ and\ \bibinfo
  {author} {\bibfnamefont {N.}~\bibnamefont {Marzari}},\ }\href@noop {}
  {\bibfield  {journal} {\bibinfo  {journal} {Physical review letters}\
  }\textbf {\bibinfo {volume} {98}},\ \bibinfo {pages} {046402} (\bibinfo
  {year} {2007})}\BibitemShut {NoStop}%
\bibitem [{\citenamefont {Sgiarovello}\ \emph {et~al.}(2001)\citenamefont
  {Sgiarovello}, \citenamefont {Peressi},\ and\ \citenamefont
  {Resta}}]{sgiarovello2001electron}%
  \BibitemOpen
  \bibfield  {author} {\bibinfo {author} {\bibfnamefont {C.}~\bibnamefont
  {Sgiarovello}}, \bibinfo {author} {\bibfnamefont {M.}~\bibnamefont
  {Peressi}},\ and\ \bibinfo {author} {\bibfnamefont {R.}~\bibnamefont
  {Resta}},\ }\href@noop {} {\bibfield  {journal} {\bibinfo  {journal}
  {Physical Review B}\ }\textbf {\bibinfo {volume} {64}},\ \bibinfo {pages}
  {115202} (\bibinfo {year} {2001})}\BibitemShut {NoStop}%
\bibitem [{\citenamefont {Marzari}\ \emph {et~al.}(2012)\citenamefont
  {Marzari}, \citenamefont {Mostofi}, \citenamefont {Yates}, \citenamefont
  {Souza},\ and\ \citenamefont {Vanderbilt}}]{marzari2012maximally}%
  \BibitemOpen
  \bibfield  {author} {\bibinfo {author} {\bibfnamefont {N.}~\bibnamefont
  {Marzari}}, \bibinfo {author} {\bibfnamefont {A.~A.}\ \bibnamefont
  {Mostofi}}, \bibinfo {author} {\bibfnamefont {J.~R.}\ \bibnamefont {Yates}},
  \bibinfo {author} {\bibfnamefont {I.}~\bibnamefont {Souza}},\ and\ \bibinfo
  {author} {\bibfnamefont {D.}~\bibnamefont {Vanderbilt}},\ }\href@noop {}
  {\bibfield  {journal} {\bibinfo  {journal} {Reviews of Modern Physics}\
  }\textbf {\bibinfo {volume} {84}},\ \bibinfo {pages} {1419} (\bibinfo {year}
  {2012})}\BibitemShut {NoStop}%
\bibitem [{\citenamefont {Pedersen}(2015)}]{pedersen2015intraband}%
  \BibitemOpen
  \bibfield  {author} {\bibinfo {author} {\bibfnamefont {T.~G.}\ \bibnamefont
  {Pedersen}},\ }\href@noop {} {\bibfield  {journal} {\bibinfo  {journal}
  {Physical Review B}\ }\textbf {\bibinfo {volume} {92}},\ \bibinfo {pages}
  {235432} (\bibinfo {year} {2015})}\BibitemShut {NoStop}%
\bibitem [{\citenamefont {Chan}\ \emph {et~al.}(2021)\citenamefont {Chan},
  \citenamefont {Qiu}, \citenamefont {da~Jornada},\ and\ \citenamefont
  {Louie}}]{chan2021giant}%
  \BibitemOpen
  \bibfield  {author} {\bibinfo {author} {\bibfnamefont {Y.-H.}\ \bibnamefont
  {Chan}}, \bibinfo {author} {\bibfnamefont {D.~Y.}\ \bibnamefont {Qiu}},
  \bibinfo {author} {\bibfnamefont {F.~H.}\ \bibnamefont {da~Jornada}},\ and\
  \bibinfo {author} {\bibfnamefont {S.~G.}\ \bibnamefont {Louie}},\ }\href@noop
  {} {\bibfield  {journal} {\bibinfo  {journal} {Proceedings of the National
  Academy of Sciences}\ }\textbf {\bibinfo {volume} {118}},\ \bibinfo {pages}
  {e1906938118} (\bibinfo {year} {2021})}\BibitemShut {NoStop}%
\bibitem [{\citenamefont {Ruan}\ \emph {et~al.}(2023)\citenamefont {Ruan},
  \citenamefont {Chan},\ and\ \citenamefont {Louie}}]{ruan2023excitonic}%
  \BibitemOpen
  \bibfield  {author} {\bibinfo {author} {\bibfnamefont {J.}~\bibnamefont
  {Ruan}}, \bibinfo {author} {\bibfnamefont {Y.-H.}\ \bibnamefont {Chan}},\
  and\ \bibinfo {author} {\bibfnamefont {S.~G.}\ \bibnamefont {Louie}},\
  }\href@noop {} {\bibfield  {journal} {\bibinfo  {journal} {arXiv preprint
  arXiv:2310.09674}\ } (\bibinfo {year} {2023})}\BibitemShut {NoStop}%
\bibitem [{\citenamefont {Morimoto}\ and\ \citenamefont
  {Nagaosa}(2016{\natexlab{b}})}]{morimoto2016exciton}%
  \BibitemOpen
  \bibfield  {author} {\bibinfo {author} {\bibfnamefont {T.}~\bibnamefont
  {Morimoto}}\ and\ \bibinfo {author} {\bibfnamefont {N.}~\bibnamefont
  {Nagaosa}},\ }\href@noop {} {\bibfield  {journal} {\bibinfo  {journal}
  {Physical Review B}\ }\textbf {\bibinfo {volume} {94}},\ \bibinfo {pages}
  {035117} (\bibinfo {year} {2016}{\natexlab{b}})}\BibitemShut {NoStop}%
\bibitem [{\citenamefont {Gao}\ \emph {et~al.}(2020)\citenamefont {Gao},
  \citenamefont {Zhang},\ and\ \citenamefont {Xiao}}]{gao2020tunable}%
  \BibitemOpen
  \bibfield  {author} {\bibinfo {author} {\bibfnamefont {Y.}~\bibnamefont
  {Gao}}, \bibinfo {author} {\bibfnamefont {Y.}~\bibnamefont {Zhang}},\ and\
  \bibinfo {author} {\bibfnamefont {D.}~\bibnamefont {Xiao}},\ }\href@noop {}
  {\bibfield  {journal} {\bibinfo  {journal} {Physical Review Letters}\
  }\textbf {\bibinfo {volume} {124}},\ \bibinfo {pages} {077401} (\bibinfo
  {year} {2020})}\BibitemShut {NoStop}%
\bibitem [{\citenamefont {Edwards}\ and\ \citenamefont
  {Thouless}(1972)}]{edwards1972numerical}%
  \BibitemOpen
  \bibfield  {author} {\bibinfo {author} {\bibfnamefont {J.}~\bibnamefont
  {Edwards}}\ and\ \bibinfo {author} {\bibfnamefont {D.}~\bibnamefont
  {Thouless}},\ }\href@noop {} {\bibfield  {journal} {\bibinfo  {journal}
  {Journal of Physics C: Solid State Physics}\ }\textbf {\bibinfo {volume}
  {5}},\ \bibinfo {pages} {807} (\bibinfo {year} {1972})}\BibitemShut {NoStop}%
\bibitem [{\citenamefont {Thouless}(1974)}]{thouless1974electrons}%
  \BibitemOpen
  \bibfield  {author} {\bibinfo {author} {\bibfnamefont {D.~J.}\ \bibnamefont
  {Thouless}},\ }\href@noop {} {\bibfield  {journal} {\bibinfo  {journal}
  {Physics Reports}\ }\textbf {\bibinfo {volume} {13}},\ \bibinfo {pages} {93}
  (\bibinfo {year} {1974})}\BibitemShut {NoStop}%
\bibitem [{\citenamefont {Lee}\ and\ \citenamefont
  {Ramakrishnan}(1985)}]{lee1985disordered}%
  \BibitemOpen
  \bibfield  {author} {\bibinfo {author} {\bibfnamefont {P.~A.}\ \bibnamefont
  {Lee}}\ and\ \bibinfo {author} {\bibfnamefont {T.~V.}\ \bibnamefont
  {Ramakrishnan}},\ }\href@noop {} {\bibfield  {journal} {\bibinfo  {journal}
  {Reviews of modern physics}\ }\textbf {\bibinfo {volume} {57}},\ \bibinfo
  {pages} {287} (\bibinfo {year} {1985})}\BibitemShut {NoStop}%
\bibitem [{\citenamefont {Shi}\ \emph {et~al.}(2021)\citenamefont {Shi},
  \citenamefont {Zhang}, \citenamefont {Chang},\ and\ \citenamefont
  {Song}}]{shi2021geometric}%
  \BibitemOpen
  \bibfield  {author} {\bibinfo {author} {\bibfnamefont {L.-k.}\ \bibnamefont
  {Shi}}, \bibinfo {author} {\bibfnamefont {D.}~\bibnamefont {Zhang}}, \bibinfo
  {author} {\bibfnamefont {K.}~\bibnamefont {Chang}},\ and\ \bibinfo {author}
  {\bibfnamefont {J.~C.}\ \bibnamefont {Song}},\ }\href@noop {} {\bibfield
  {journal} {\bibinfo  {journal} {Physical Review Letters}\ }\textbf {\bibinfo
  {volume} {126}},\ \bibinfo {pages} {197402} (\bibinfo {year}
  {2021})}\BibitemShut {NoStop}%
\bibitem [{\citenamefont {Taherinejad}\ \emph {et~al.}(2014)\citenamefont
  {Taherinejad}, \citenamefont {Garrity},\ and\ \citenamefont
  {Vanderbilt}}]{taherinejad2014wannier}%
  \BibitemOpen
  \bibfield  {author} {\bibinfo {author} {\bibfnamefont {M.}~\bibnamefont
  {Taherinejad}}, \bibinfo {author} {\bibfnamefont {K.~F.}\ \bibnamefont
  {Garrity}},\ and\ \bibinfo {author} {\bibfnamefont {D.}~\bibnamefont
  {Vanderbilt}},\ }\href@noop {} {\bibfield  {journal} {\bibinfo  {journal}
  {Physical Review B}\ }\textbf {\bibinfo {volume} {89}},\ \bibinfo {pages}
  {115102} (\bibinfo {year} {2014})}\BibitemShut {NoStop}%
\bibitem [{\citenamefont {Gresch}\ \emph {et~al.}(2017)\citenamefont {Gresch},
  \citenamefont {Autes}, \citenamefont {Yazyev}, \citenamefont {Troyer},
  \citenamefont {Vanderbilt}, \citenamefont {Bernevig},\ and\ \citenamefont
  {Soluyanov}}]{gresch2017z2pack}%
  \BibitemOpen
  \bibfield  {author} {\bibinfo {author} {\bibfnamefont {D.}~\bibnamefont
  {Gresch}}, \bibinfo {author} {\bibfnamefont {G.}~\bibnamefont {Autes}},
  \bibinfo {author} {\bibfnamefont {O.~V.}\ \bibnamefont {Yazyev}}, \bibinfo
  {author} {\bibfnamefont {M.}~\bibnamefont {Troyer}}, \bibinfo {author}
  {\bibfnamefont {D.}~\bibnamefont {Vanderbilt}}, \bibinfo {author}
  {\bibfnamefont {B.~A.}\ \bibnamefont {Bernevig}},\ and\ \bibinfo {author}
  {\bibfnamefont {A.~A.}\ \bibnamefont {Soluyanov}},\ }\href@noop {} {\bibfield
   {journal} {\bibinfo  {journal} {Physical Review B}\ }\textbf {\bibinfo
  {volume} {95}},\ \bibinfo {pages} {075146} (\bibinfo {year}
  {2017})}\BibitemShut {NoStop}%
\bibitem [{\citenamefont {Fu}\ and\ \citenamefont {Kane}(2006)}]{fu2006time}%
  \BibitemOpen
  \bibfield  {author} {\bibinfo {author} {\bibfnamefont {L.}~\bibnamefont
  {Fu}}\ and\ \bibinfo {author} {\bibfnamefont {C.~L.}\ \bibnamefont {Kane}},\
  }\href@noop {} {\bibfield  {journal} {\bibinfo  {journal} {Physical Review
  B—Condensed Matter and Materials Physics}\ }\textbf {\bibinfo {volume}
  {74}},\ \bibinfo {pages} {195312} (\bibinfo {year} {2006})}\BibitemShut
  {NoStop}%
\end{thebibliography}%

\clearpage

\setcounter{equation}{0}               
\renewcommand{\theequation}{S\arabic{equation}}

\begin{widetext}

\section*{Supplementary Information for ``Correlated Quantum Shift Vector of Particle-Hole Excitations''}

\section{Bethe-Salpeter equation in relative coordinates}
It is instructive to express the Bethe-Salpeter equation (BSE) in terms of relative coordinates. Below we use $\textbf{R}$ to denote the real coordinates of individual electrons/holes, $\textbf{R}_{\rm cm}$ to denote the center-of-mass coordinates of electron-hole pair, and $\textbf{r}$ to denote the relative coordinates between electrons and holes.

We use $c^{\dagger}$ to denote creation operators for Wannier functions, which can be represented by \begin{align}
    &\ket{n,\textbf{R}}\equiv c^{\dagger}_{n,\textbf{R}}\ket{0}=\int d\textbf{x }w_{n,\textbf{R}}(\textbf{x})a^{\dagger}_{\textbf{x}}\ket{0} ,
\end{align}
where $n$ is the band index and $a_{\textbf{x}}$ is the field operator in free space. Here $w_{n,\textbf{R}}(\textbf{x})$ is the exponentially localized Wannier functions centered around $\textbf{R}$ which decays as $e^{-|\textbf{x}-\textbf{R}|/\xi_W}$ at large $|\textbf{x}-\textbf{R}|$ and $\xi_W$ is the Wannier function extent. Using the completeness relation $\sum\limits_{n,\textbf{R}}w_{n,\textbf{R}}(\textbf{x})w^*_{n,\textbf{R}}(\textbf{x}')=\delta(\textbf{x}-\textbf{x}')$, we can expand $a^{\dagger}_{\textbf{x}}$ in terms of $c^{\dagger}_{n,\textbf{R}}$, where  $n$ is the band index:
\begin{align}
    &a^{\dagger}_{\textbf{x}}\ket{0}=\sum\limits_{n,\textbf{R}}w_{n,\textbf{R}}^*(\textbf{x})c^{\dagger}_{n,\textbf{R}}\ket{0}.
\end{align}

Starting from the eigen equation $\hat{H}\ket{\psi_{\bm{Q}}}_{\text{p}-\text{h}}=E\ket{\psi_{\bm{Q}}}_{\text{p}-\text{h}}$, we can project it onto the particle-hole basis $\ket{\textbf{R}_{\rm cm},\textbf{r}}$ to arrive at the BSE for envelope functions: $\sum\limits_{\textbf{r}'}\mathcal{H}_{\bm{Q}}(\textbf{r},\textbf{r}')\psi_{\bm{Q}}(\textbf{r}')=E\psi_{\bm{Q}}(\textbf{r})$. The BSE 
Hamitonian is therefore:
\begin{align}
&\mathcal{H}_{\bm{Q}}(\textbf{r},\textbf{r}')=\frac{1}{N}\sum\limits_{\textbf{R}_{\rm cm,1},\textbf{R}_{\rm cm,2}}e^{i\textbf{Q}\cdot(\textbf{R}_{\rm cm,2}-\textbf{R}_{\rm cm,1})}\bra{\textbf{R}_{\rm cm,1},\textbf{r}}\hat{H}\ket{\textbf{R}_{\rm cm,2},\textbf{r}'}
\end{align}

\textbf{Kinetic term}: the kinetic term $\mathcal{H}_{\textbf{K}}(\textbf{r},\textbf{r}')$ of BSE Hamiltonian is obtained by projecting the kinetic part $\hat{H}_{\textbf{K}}$ of the total Hamiltonian onto the particle-hole basis, where $\hat{H}_{\textbf{K}}=\int d\textbf{x} a_{\textbf{x}}^{\dagger}(\frac{\hat{\textbf{p}}^2}{2m}+V_{\rm lattice}(\textbf{x}))a_{\textbf{x}}$. After a straightforward evaluation, we arrive at:
\begin{align}
&\mathcal{H}_{\textbf{K}}(\textbf{r},\textbf{r}')=e^{i\textbf{Q}\cdot (\textbf{r}'-\textbf{r})/2}\bra{c,\textbf{r}}\hat{H}_{\textbf{K}}\ket{c,\textbf{r}'}-e^{i\textbf{Q}\cdot(\textbf{r}-\textbf{r}')/2}\bra{v,\textbf{r}}\hat{H}_{\textbf{K}}\ket{v,\textbf{r}'},
\end{align}
where $\bra{c,\textbf{r}}\hat{H}_{\textbf{K}}\ket{c,\textbf{r}'}\equiv \int d\textbf{x } w_{c,\textbf{r}}^*(\textbf{x})(\frac{\hat{\textbf{p}}^2}{2m}+V_{\rm lattice}(\textbf{x}))w_{c,\textbf{r}'}(\textbf{x})$, 
which is of order $e^{-|\textbf{r}-\textbf{r}'|/\xi_W}$ from the exponential localization nature of Wannier functions (and similarly for $v$ bands). 

Utilizing the relation between the Bloch functions and the Wannier functions $w_{\textbf{R}}(\textbf{x})=\frac{1}{\sqrt{N}}\sum\limits_{\textbf{k}}e^{-i\textbf{k}\cdot\textbf{R}}\psi_{\textbf{k}}(\textbf{x})$, we can immediately see that the kinetic term $\mathcal{H}_{\textbf{K}}(\textbf{r},\textbf{r}')=\frac{1}{N}\sum\limits_{\textbf{p}}e^{i\textbf{p}\cdot(\textbf{r}-\textbf{r}')}[\epsilon_c(\textbf{p}+\textbf{Q}/2)-\epsilon_v(\textbf{p}-\textbf{Q}/2)]$ as discussed in the main text.

\textbf{Interaction term}: the interaction part of the Hamiltonian is of the general form $\hat{H}_{\textbf{V}}=\int d\textbf{x}_1d\textbf{x}_2V(|\textbf{x}_1-\textbf{x}_2|)a^{\dagger}_{\textbf{x}_1}a^{\dagger}_{\textbf{x}_2}a_{\textbf{x}_2}a_{\textbf{x}_1}$. The electron-electron interaction 
$V(|\textbf{x}|)$ can either be screened or un-screened and vanishes as $|\textbf{x}|\rightarrow \infty$. After projecting onto the particle-hole basis and performing Hartree-Fock contractions, we arrive at two terms, the direct term $\mathcal{V}_{D}$ and the exchange interaction term $\mathcal{V}_{X}$.

The direct interaction term:
\begin{align}\label{eq:appendix_Vdirect}
  & \mathcal{V}_{D}(\textbf{r},\textbf{r}')=-\sum\limits_{\Delta\textbf{R}_{\rm cm}}e^{i\textbf{Q}\cdot(\Delta\textbf{R}_{\rm cm})}\int d\textbf{x}_1d\textbf{x}_2 w_{c,\Delta\textbf{R}_{\rm cm}+\textbf{r}'/2}(\textbf{x}_1)w^*_{c,\textbf{r}/2}(\textbf{x}_1)V(|\textbf{x}_1-\textbf{x}_2|)w_{v,-\textbf{r}/2}(\textbf{x}_2)w^*_{v,\Delta\textbf{R}_{\rm cm}-\textbf{r}'/2}(\textbf{x}_2),
\end{align}
where the minus sign comes from electron anti-commutations and $\Delta\textbf{R}_{\rm cm}=\textbf{R}_{\rm cm,2}-\textbf{R}_{\rm cm,1}$.

Physically, each coordinate $\textbf{x}_i$ only takes significant non-zero values near the location of the Wannier functions which makes the $\mathcal{V}_D(\textbf{r},\textbf{r}')$ almost diagonal in the relative coordinate space. A simple estimate can be made by noticing that e.g., $w_{n,\textbf{R}}(\textbf{x})$ is upper bounded by $e^{-|\textbf{R}-\textbf{x}|/\xi_W}$. Therefore we have
\begin{equation}
    \begin{split}
    &|\mathcal{V}_D(\textbf{r},\textbf{r}')|<e^{-(|\Delta\textbf{R}_{\rm cm}+\textbf{r}'/2-\textbf{x}_1|+|\textbf{r}/2-\textbf{x}_1|+|-\textbf{r}/2-\textbf{x}_2|+|\Delta\textbf{R}_{\rm cm}-\textbf{r}'/2-\textbf{x}_2|)/\xi_W}\\
    &<e^{-(|(\Delta\textbf{R}_{\rm cm}+\textbf{r}'/2-\textbf{x}_1)-(\textbf{r}/2-\textbf{x}_1)+(-\textbf{r}/2-\textbf{x}_2)-(\Delta\textbf{R}_{\rm cm}-\textbf{r}'/2-\textbf{x}_2)|)/\xi_W}=e^{-(|\textbf{r}'-\textbf{r}|)/\xi_W}.        
    \end{split}
\end{equation}
This means that $\mathcal{V}_D(\textbf{r},\textbf{r}')$ is non-zero only when $\textbf{r}\sim\textbf{r}'$ (i.e., nearly diagonal in $\textbf{r}$). 

The exchange interaction is:
\begin{align}\label{eq:appendix_Vexchange}
  & \mathcal{V}_{X}(\textbf{r},\textbf{r}')=\sum\limits_{\Delta\textbf{R}_{\rm cm}}e^{i\textbf{Q}\cdot(\Delta\textbf{R}_{\rm cm})}\int d\textbf{x}_1d\textbf{x}_2 w_{c,\Delta\textbf{R}_{\rm cm}+\textbf{r}'/2}(\textbf{x}_1)w^*_{v,\Delta\textbf{R}_{\rm cm}-\textbf{r}'/2}(\textbf{x}_1)V(|\textbf{x}_1-\textbf{x}_2|)w_{v,-\textbf{r}/2}(\textbf{x}_2)w^*_{c,\textbf{r}/2}(\textbf{x}_2),
\end{align}

Since it is the exchange energy of \textbf{electron} and \textbf{hole}, the electron and hole should be close to have non negligible exchange (meaning that the relative coordinate should be small). For example if we only look at the product of Wannier functions with common coordinate $\textbf{x}_1$, then it is upper-bounded by:
\begin{align}
    &e^{-(|\Delta\textbf{R}_{cm}+\textbf{r}'/2-\textbf{x}_1|+|\Delta\textbf{R}_{\rm cm}-\textbf{r}'/2-\textbf{x}_1|)/\xi_W}<e^{-(|(\Delta\textbf{R}_{\rm cm}+\textbf{r}'/2-\textbf{x}_1)-(\Delta\textbf{R}_{\rm cm}-\textbf{r}'/2-\textbf{x}_1)|)/\xi_W}=e^{-|\textbf{r}'|/\xi_W}.
\end{align}
Similarly the product of Wannier functions with common coordinate $\textbf{x}_2$ is upper-bounded by $e^{-|\textbf{r}|/\xi_W}$:
\begin{align}
    &e^{-(|-\textbf{r}/2-\textbf{x}_2|+|\textbf{r}/2-\textbf{x}_2|)/\xi_W}<e^{-(|(-\textbf{r}/2-\textbf{x}_2)-(\textbf{r}/2-\textbf{x}_2)|)/\xi_W}=e^{-|\textbf{r}|/\xi_W}.
\end{align}

As a result, the full $\mathcal{V}_X(\textbf{r},\textbf{r}')$ is therefore upper bounded by $e^{-(|\textbf{r}|+|\textbf{r}'|)/\xi_W}$, meaning that it only concentrates near $|\textbf{r}|,|\textbf{r}'|\sim 0$. Furthermore we have $\mathcal{V}_X(\textbf{r},\textbf{r}')<e^{-(|\textbf{r}|+|\textbf{r}'|)/\xi_W}<e^{-|\textbf{r}-\textbf{r}'|/\xi_W}$, meaning $\mathcal{V}_X$ is still almost diagonal.

In summary, we have given the explicit form of Bethe-Salpeter Hamiltonian in relative coordinates as the sum of three terms $\mathcal{H}\equiv \mathcal{H}_{\textbf{K}}+\mathcal{V}_D+\mathcal{V}_X$ and verified the exponentially decaying $e^{-|\textbf{r}-\textbf{r}'|/\xi_W}$ of the off-diagonal elements $\mathcal{H}(\textbf{r},\textbf{r}')$.

\section{Flux insertion and the transformation rule of matrix elements}
Following Kohn\cite{kohn1964theory,souza2000polarization}, we introduce the following parameterized Hamiltonian on a periodic lattice with linear dimension $L$:  
\begin{equation}
   \hat{H}(\bm{\kappa})=\frac{1}{2m}\sum\limits_{i=1}^N(\hat{\textbf{p}}_i+\hbar \bm{\kappa})^2+\hat{V}_{\rm lattice}+\hat{H}_{\rm int},
\end{equation}
where $\bm{\kappa}$ is a uniform flux (units inverse length), $\hat{V}_{\rm lattice}$ is the lattice potential and $\hat{H}_{\rm int}$ describes electron-electron interactions. We label the $n$-th eigen-states of $\hat{H}(\bm{\kappa})$ as $\ket{\Phi_n(\bm{\kappa})}$. The vector $\bm{\kappa}$ can be interpreted as arising from the insertion of a flux through the ring formed along the periodic direction. Henceforth it will be referred to as a flux insertion or twisted boundary condition according to literature. 

From the free part (the sum of kinetic energy and $\hat{V}_{\rm lattice}$), we can solve for the Bloch functions and then Fourier transforming them to Wannier functions. Below we will 
track the change of Wannier functions and the matrix elements of BSE Hamiltonian in the Wannier basis in response to flux insertion. As discussed in Ref.~\cite{luttinger1951effect,kohn1964theory}, the Wannier function corresponding to the free part after flux insertion is related to that before flux insertion via $w_{n,\textbf{R}}^{\bm \kappa}(\textbf{x})=e^{-i\bm{\kappa}\cdot(\textbf{x}-\textbf{R})}{w}_{n,\textbf{R}}(\textbf{x}).$ We shall denote the corresponding creation operator as $c^{\dagger}_{\bm{\kappa};n,\textbf{R}}$.

We now turn to analyzing the Hamiltonian terms under flux insertion. Any general particle-conserving one-body operator $\hat{G}\equiv \int d\textbf{x} a^{\dagger}_{\textbf{x}}G(\hat{\textbf{x}},\hat{\textbf{p}})a_{\textbf{x}}$ can be expressed in the Wannier basis as: 
\begin{align}
\hat{G}=\sum\limits_{n,m,\textbf{R},\textbf{R}'}\bra{n,\textbf{R}}\hat{G}\ket{m,\textbf{R}'}c^{\dagger}_{n,\textbf{R}}c_{m,\textbf{R}'},
\end{align}
with $\bra{n,\textbf{R}}\hat{G}\ket{m,\textbf{R}'}\equiv \int d\textbf{x}w^*_{n,\textbf{R}}(\textbf{x})G(\hat{\textbf{x}},\hat{\textbf{p}})w_{m,\textbf{R}'}(\textbf{x})$.

Upon inserting a uniform flux, the operator becomes: $\hat{G}^{\bm{\kappa}}= \int d\textbf{x} a^{\dagger}_{\textbf{x}}G(\hat{\textbf{x}},\hat{\textbf{p}}+\hbar\bm{\kappa})a_{\textbf{x}}$, and the matrix elements in the flux-inserted Wannier basis $\ket{\bm{\kappa};n,\textbf{R}}\equiv c^{\dagger}_{\bm{\kappa};n,\textbf{R}}\ket{0}$ are given by:
\begin{align}
&\int d\textbf{x}w^*_{n,\textbf{R}}(\textbf{x})e^{i\bm{\kappa}\cdot(\textbf{x}-\textbf{R})}G(\hat{\textbf{x}},\hat{\textbf{p}}+\hbar\bm{\kappa})e^{-i\bm{\kappa}\cdot(\textbf{x}-\textbf{R}')}w_{m,\textbf{R}'}(\textbf{x})=e^{-i\bm{\kappa}\cdot(\textbf{R}-\textbf{R}')}\int d\textbf{x}w^*_{n,\textbf{R}}(\textbf{x})G(\hat{\textbf{x}},\hat{\textbf{p}})w_{m,\textbf{R}'}(\textbf{x}),
\end{align}
where we have used $e^{i\bm{\kappa}\cdot\textbf{x}}G(\hat{\textbf{x}},\hat{\textbf{p}}+\hbar\bm{\kappa})e^{-i\bm{\kappa}\cdot\textbf{x}}=G(\hat{\textbf{x}},\hat{\textbf{p}})$ recalling that $(\hat{\textbf{p}} + \hbar \bm \kappa)e^{-i\bm \kappa \cdot \textbf{x}} = e^{-i\bm \kappa \cdot \textbf{x}} \hat{\textbf{p}}$. 

Thus, the full operator after flux insertion becomes : 
\begin{align}
\hat{G}^{\bm{\kappa}}=\sum\limits_{n,m,\textbf{R},\textbf{R}'}e^{-i\bm{\kappa}\cdot(\textbf{R}-\textbf{R}')}\bra{n,\textbf{R}}\hat{G}\ket{m,\textbf{R}'}c^{\dagger}_{\bm{\kappa};n,\textbf{R}}c_{\bm{\kappa};m,\textbf{R}'}
\end{align}

Similarly, for any general particle-conserving two-body operator $\hat{F}\equiv\int d\textbf{x}_1\int d\textbf{x}_2 a^{\dagger}_{\textbf{x}_1}a^{\dagger}_{\textbf{x}_2}F(\hat{\textbf{x}}_1,\hat{\textbf{x}}_2)a_{\textbf{x}_2}a_{\textbf{x}_1},$ the matrix element $\bra{\textbf{R}_1,\textbf{R}_2}\hat{F}\ket{\textbf{R}_3,\textbf{R}_4}$ in the Wannier basis transforms under flux insertion as:
\begin{equation}
\bra{\textbf{R}_1,\textbf{R}_2}\hat{F}\ket{\textbf{R}_3,\textbf{R}_4}\rightarrow e^{-i\bm{\kappa}\cdot(\textbf{R}_1+\textbf{R}_2-\textbf{R}_3-\textbf{R}_4)}\bra{\textbf{R}_1,\textbf{R}_2}\hat{F}\ket{\textbf{R}_3,\textbf{R}_4}. 
\end{equation}

For the BSE Hamiltonian, by directly substituting the flux-threaded Wannier functions into each of its terms, namely, $\mathcal{H}_{\textbf{K}}$, $\mathcal{V}_{D}$, $\mathcal{V}_{X}$, we obtain the following transformation rule of matrix elements $\mathcal{H}(\textbf{r},\textbf{r}')$ under flux insertion:
\begin{align}
   &\mathcal{H}^{\bm{\kappa}}(\textbf{r},\textbf{r}')= e^{-i\bm{\kappa}\cdot (\textbf{r}-\textbf{r}')}\mathcal{H}(\textbf{r},\textbf{r}'). 
\end{align}
where we have noted the opposite charge for electrons and holes in the BSE mean that fluxes accumulate through the relative coordinate $\textbf{r}$.

\section{deviation estimation of the ansatz function for exciton states with flux insertion}
Let's first discuss how periodic boundary condition is implemented in BSE. To impose periodic boundary conditions, we restrict coordinates to $-L/2 < r_x < L/2$ and take $\bm{\kappa}$ along the $x$ direction. On a periodic lattice, the Bethe–Salpeter Hamiltonian is defined with the minimal distance:
\begin{align}
    \mathcal{H}(\textbf{r},\textbf{r}')\equiv \mathcal{H}(\textbf{r},\textbf{r}'+ML\textbf{e}_x), \text{ with }M \in \mathbb{Z} \text{  minimizing  } |r_x - r_x' - ML|.
\end{align}
Correspondingly, flux insertion modifies matrix elements by
\begin{align}
    \mathcal{H}^{\bm{\kappa}}(\textbf{r},\textbf{r}')=e^{-ik_x(r_x-r_x'-ML)}\mathcal{H}(\textbf{r},\textbf{r}')
\end{align}

For instance, when $r_x \lesssim L/2$ and $r_x' \gtrsim -L/2$, the minimal distance is $r_x - r_x' - L$ with $M = 1$.

We now proceed to estimate the deviation of the ansatz $\tilde{\psi}^{\bm{\kappa}}(\textbf{r}) = e^{-i\bm{\kappa} \cdot \textbf{r}} \psi(\textbf{r})$ for the exciton envelope function, we consider a bound state solution $\psi(\textbf{r})$ of the BSE at $\bm{\kappa}=0$ satisfying $\sum_{\bm r'} \mathcal{H} (\bm r,\bm r') \psi (\bm r') = E\psi (\bm r)$ and exponential decaying $|\psi(\textbf{r})| \sim e^{-|\textbf{r}|/\xi}$. Here and in what follows, we have suppressed the subscript $\bm Q$ for the center-of-mass momentum 
as it is conserved and decoupled from $\bm{\kappa}$. As we will see below, the key point is that the ansatz $\tilde{\psi}^{\bm{\kappa}}$ only violates the BSE near the boundaries $r_x \sim \pm L/2$, where the wavefunction is exponentially small. Thus, the deviation is exponentially suppressed in system size, making the ansatz asymptotically exact. 

To see this explicitly, we quantify the deviation of the ansatz $\tilde{\psi}^{\bm{\kappa}}(\textbf{r})$ by examining the following deviation:
\begin{align}
\label{eq:SuppError}
\zeta(\textbf{r})\equiv\left| \sum_{\textbf{r}'} \mathcal{H}^{\bm{\kappa}}(\textbf{r}, \textbf{r}') \tilde{\psi}^{\bm{\kappa}}(\textbf{r}') - E \tilde{\psi}^{\bm{\kappa}}(\textbf{r}) \right|.
\end{align}
Without loss of generality, we assume $0 < r_x < L/2$. To evaluate $\zeta(\textbf{r})$, we split the sum over $\textbf{r}'$ into a \textit{bulk region} $\mathbb{R}_{\text{bulk}}$ where $|r_x - r_x'| < L/2$, and an \textit{edge region} $\mathbb{R}_{\text{edge}}$ where $|r_x - r_x'| > L/2$. The Hamiltonian matrix elements relate to those without flux as:
\[
\mathcal{H}^{\bm{\kappa}}(\textbf{r}, \textbf{r}') = 
\begin{cases}
e^{-i\bm{\kappa} \cdot (\textbf{r} - \textbf{r}')} \mathcal{H}(\textbf{r}, \textbf{r}'), & \textbf{r}' \in \mathbb{R}_{\text{bulk}}, \\
e^{-i\bm{\kappa} \cdot (\textbf{r} - \textbf{r}' - L \textbf{e}_x)} \mathcal{H}(\textbf{r}, \textbf{r}'), & \textbf{r}' \in \mathbb{R}_{\text{edge}}.
\end{cases}
\]

Inserting this into the flux-threaded equation yields:
\begin{align}
\sum_{\textbf{r}'} \mathcal{H}^{\bm{\kappa}}(\textbf{r}, \textbf{r}') \tilde{\psi}^{\bm{\kappa}}(\textbf{r}')
= e^{-i\bm{\kappa}\cdot \textbf{r}} \left[ \sum_{\mathbb{R}_{\text{bulk}}} \mathcal{H}(\textbf{r}, \textbf{r}') \psi(\textbf{r}') 
+ e^{i\kappa_x L} \sum_{\mathbb{R}_{\text{edge}}} \mathcal{H}(\textbf{r}, \textbf{r}') \psi(\textbf{r}') \right].
\end{align}
Subtracting $E\tilde{\psi}^{\bm{\kappa}} (\bm r)= e^{-i\bm{\kappa}\cdot \textbf{r}} E\psi(\textbf{r})$, we find that the deviation in Eq.~(\ref{eq:SuppError}) manifests in $\mathbb{R}_{\rm edge}$:
\begin{align}\label{appendix:error}
\zeta(\textbf{r})= \left| e^{-i\bm{\kappa} \cdot \textbf{r}} (e^{i\kappa_x L} - 1) \sum_{\mathbb{R}_{\text{edge}}} \mathcal{H}(\textbf{r}, \textbf{r}') \psi(\textbf{r}') \right|.
\end{align}

For localized exciton envelope functions, $\psi(\textbf{r}) \sim e^{-|\textbf{r}|/\xi}$ and $\mathcal{H}(\textbf{r}, \textbf{r}')$ decays with range $\xi_W$. Bounding the sum: 
\begin{align}
&\zeta(\textbf{r}) <|e^{i\kappa_xL}-1|\cdot e^{-|\textbf{r}-\textbf{r}'-L\textbf{e}_x|/\xi_W}\cdot e^{-|\textbf{r}'|/\xi}<|e^{i\kappa_xL}-1|\cdot e^{-|\textbf{r}-L\textbf{e}_x|/\xi_{M}}<|e^{i\kappa_xL}-1|\cdot e^{-L/(2\xi_{M})}
\end{align}
where $\xi_M = \max(\xi, \xi_W)$ and we used the triangular inequality and $0 < r_x < L/2$. This displays that deviation is exponentially suppressed.

As a result,  we arrive at 
\begin{align}
\sum_{\textbf{r}'} \mathcal{H}^{\bm{\kappa}}(\textbf{r}, \textbf{r}') \tilde{\psi}^{\bm{\kappa}}(\textbf{r}') 
= E \tilde{\psi}^{\bm{\kappa}}(\textbf{r}) + \mathcal{O}(e^{-L / (2\xi_M)}),
\end{align}
or equivalently,
\begin{align}
\psi^{\bm{\kappa}}(\textbf{r}) = e^{-i\bm{\kappa} \cdot \textbf{r}} \psi(\textbf{r}) + \mathcal{O}(e^{-L / (2\xi_M)}),\qquad
E(\bm{\kappa}) = E(0) + \mathcal{O}(e^{-L / \xi_M}),
\end{align}
i.e., $\tilde{\psi}^{\bm{\kappa}}$ provides an exponentially accurate approximation to the true eigenstate $\psi^{\bm{\kappa}}$, with an energy shift that is exponentially suppressed. This mirrors the Thouless criterion for localized states vs delocalized states \cite{edwards1972numerical,thouless1974electrons,lee1985disordered}.


In contrast, note that 
Eq.~\eqref{appendix:error} remains valid for as for extended states such as plane waves envelopes ($|\psi(\textbf{r})| \sim L^{-d/2}$), when the form of the delocalized envelope for free particle-hole excitation is substituted, significant deviations occur. 
Indeed, when $r_x \sim L/2$ and $r_x' \sim -L/2$, the amplitude $|\psi(\textbf{r}')| \sim L^{-d/2}$ leads to a sizable deviation:
\begin{align}
\zeta(\textbf{r}) \sim |e^{i\kappa_x L} - 1| \cdot \mathcal{O}(L^{-d/2}),
\end{align}
which constitutes a non-negligible correction for an extended envelope function. Thus, for extended states the ansatz $\tilde{\psi}_{\bm{\kappa}}$ fails to solve the flux-threaded Bethe–Salpeter equation. This 
mirrors 
the breakdown of ``gauge invariance” for extended states in a periodic system~\cite{kohn1964theory}.

\section{Insensitivity of exciton transition shift vector vs. $\hat{V}$}
In this section, we first prove in general that the exciton transition shift vector is insensitive to the form of an interaction $\hat{V}$ inducing the transition from ground state to exciton state; we then extend it to show that transitions between excitonic states are also insensitive to the form of $\hat{V}$. To that end, we consider exciton shift vectors induced by two general particle-conserving interactions $\hat{V}_1$ and $\hat{V}_2$ and evaluate their differences $\delta \bm{\mathcal{R}} = \bm{\mathcal{R}}^{V_1}_{0\rightarrow {\rm ex}}-\bm{\mathcal{R}}^{V_2}_{0\rightarrow {\rm ex}}$, which amounts to proving Eq.~\eqref{eq:shift_vector_difference} of the main text.

For any particle-conserving one-body interaction $\hat{V}_1$, we can expand them in the Wannier basis as $\hat{V}_1\equiv \sum\limits_{\textbf{R},\textbf{R}'}(V_{1})_{\textbf{R},\textbf{R}'}^{n,m}c^{\dagger}_{n,\textbf{R}}c_{m,\textbf{R}'}$, where $n,m$ are band indices. (The treatment of particle-conserving two-body interaction is exactly the same.) Following the previous discussion, the matrix elements transform as $(V_1)^{n,m}_{\textbf{R},\textbf{R}'}\rightarrow e^{-i\bm{\kappa}\cdot (\textbf{R}-\textbf{R}')}(V_1)^{n,m}_{\textbf{R},\textbf{R}'}$ under flux insertion. The transformation rule holds only for particle-conserving interactions and is violated in particle-non-conserving interactions such as $c^{\dagger}c^{\dagger}$, as expected in a superconductor. 

The matrix element of $\bra{\Phi_0(\bm{\kappa})}\hat{V}_1\ket{\Phi_{\text{ex}}(\bm{\kappa})}$ can then be evaluated with Wick's theorem and yields $\sum\limits_{\textbf{R},\textbf{R}'}e^{i\textbf{Q}\cdot (\textbf{R}+\textbf{R}')/2}(V_1)_{\textbf{R},\textbf{R}'}^{v,c}\psi_{\textbf{Q}}(\textbf{R}'-\textbf{R})$, which is independent of $\bm{\kappa}$ (with a deviation $\mathcal{O}(e^{-L/(2\xi_M)})$ from the envelope function) as the phase change of $(V_1)^{v,c}_{\textbf{R},\textbf{R}'}$ and $\psi_{\textbf{Q}}(\textbf{R}'-\textbf{R})$ arising from flux insertion exactly cancels. Similarly we can show that $\bra{\Phi_{\text{ex}}(\bm{\kappa})}\hat{V}_2\ket{\Phi_0(\bm{\kappa})}$ is also independent of $\bm{\kappa}$ also with a deviation $\mathcal{O}(e^{-L/(2\xi_M)})$. Although each term $\bra{\Phi_0(\bm{\kappa})}\hat{V}_1\ket{\Phi_{\text{ex}}(\bm{\kappa})}$ and $\bra{\Phi_{\text{ex}}(\bm{\kappa})}\hat{V}_2\ket{\Phi_0(\bm{\kappa})}$ are both gauge dependent, their product is gauge invariant and is independent of $\bm{\kappa}$ with a combined deviation $\mathcal{O}(e^{-L/\xi_M})$. Therefore we arrive at Eq.~\eqref{eq:shift_vector_difference} shown in the main text:
\begin{equation}
\begin{split}
&\bm{\mathcal{R}}^{V_1}_{0\rightarrow \text{ex}}-\bm{\mathcal{R}}^{V_2}_{0\rightarrow \text{ex}}=\nabla_{\bm{\kappa}}\text{arg}(\bra{\Phi_0({\bm{\kappa}})}\hat{V}_1\ket{\Phi_{\text{ex}}({\bm{\kappa}})}\bra{\Phi_{\text{ex}}({\bm{\kappa}})}\hat{V}_2\ket{\Phi_0({\bm{\kappa}})})=\mathcal{O}(e^{-L/\xi_M}).
\end{split}
\end{equation}

Furthermore, we can establish the following relation for the shift vector between two excitons ex$_1$ and ex$_2$: 
\begin{align}
\bm{\mathcal{R}}^{V_{1 2}}_{\text{ex}_1\rightarrow \text{ex}_2}\doteq\bm{\mathcal{R}}^{V_{02}}_{0\rightarrow \text{ex}_2}-\bm{\mathcal{R}}^{V_{01}}_{0\rightarrow \text{ex}_1}. 
\end{align}
This reduces to showing that $\nabla_{\bm{\kappa}}\text{arg}(\bra{\Phi_0({\bm{\kappa}})}\hat{V}_{01}\ket{\Phi_{\text{ex}_1}({\bm{\kappa}})}\bra{\Phi_{\text{ex}_1}({\bm{\kappa}})}\hat{V}_{12}\ket{\Phi_{\text{ex}_2}({\bm{\kappa}})}\bra{\Phi_{\text{ex}_2}({\bm{\kappa}})}\hat{V}_{02}\ket{\Phi_0({\bm{\kappa}})})\doteq 0$, which can be demonstrated in the same manner as before, with a deviation of order $\mathcal{O}(e^{-2L/\xi_M})$, arising from the four appearances of the exciton envelope functions in the full Wilson loop. 

\section{Intrinsic expression for exciton transition shift vector}
We evaluate the exciton shift vector using the exciton wave function $\ket{\psi_{\bm{Q}}}=\frac{1}{\sqrt{N}}\sum\limits_{\textbf{r},\textbf{R}_{\rm cm}}\psi_{\bm{Q}}(\textbf{r})e^{i\bm{Q}\cdot\textbf{R}_{\rm cm}}\ket{\textbf{R}_{\rm cm},\textbf{r}}$ along with the definition of the shift vector:
\begin{align}
\bm{\mathcal{R}}_{0n}\equiv i\bra{\Phi_n}\nabla_{\bm{\kappa}}\ket{\Phi_n}-i\bra{\Phi_0}\nabla_{\bm{\kappa}}\ket{\Phi_0}+\nabla_{\bm{\kappa}}\text{arg}(\bra{\Phi_0}\hat{V}\ket{\Phi_n}).
\end{align}

Under flux insertion, the Wannier functions transform as $w_{\textbf{R}}(\textbf{x})\rightarrow w^{\bm{\kappa}}_{\textbf{R}}(\textbf{x})=e^{-i\bm{\kappa}\cdot(\textbf{x}-\textbf{R})}w_{\textbf{R}}(\bm{x})$. Let's denote the corresponding creation operator as $c^{\dagger}_{\bm{\kappa};\textbf{R}}$. 

Under flux insertion we can write the exciton state as 
\begin{align}
\ket{\psi^{\bm{\kappa}}_{\bm{Q}}}=\frac{1}{\sqrt{N}}\sum\limits_{\textbf{r},\textbf{R}_{\rm cm}}\psi^{\bm{\kappa}}_{\bm{Q}}(\textbf{r})e^{i\bm{Q}\cdot\textbf{R}_{\rm cm}}c^{\dagger}_{\bm{\kappa};c,\textbf{R}_{\rm cm}+\textbf{r}/2}c_{\bm{\kappa};v,\textbf{R}_{\rm cm}-\textbf{r}/2}\ket{\rm GS;\bm{\kappa}},
\end{align}
where $\ket{\text{GS};\bm{\kappa}}=\prod_{\textbf{R}}c^{\dagger}_{\bm{\kappa};v,\textbf{R}}\ket{0}$. Recalling that $\psi_{\bm{Q}}^{\bm{\kappa}}(\textbf{r})=e^{-i\bm{\kappa}\cdot\textbf{r}}\psi_{\bm{Q}}(\textbf{r})$, we find the exciton transition shift vector is 
%
%
\begin{align}
    \bm{\mathcal{R}}_{0\rightarrow \rm ex}\doteq i\bra{\psi^{\bm{\kappa}}_{\bm{Q}}}\nabla_{\bm{\kappa}}\big[\frac{1}{\sqrt{N}}\sum\limits_{\textbf{r},\textbf{R}_{\rm cm}}\psi^{\bm{\kappa}}_{\bm{Q}}(\textbf{r})e^{i\bm{Q}\cdot\textbf{R}_{\rm cm}}c^{\dagger}_{\bm{\kappa};c,\textbf{R}_{\rm cm}+\textbf{r}/2}c_{\bm{\kappa};v,\textbf{R}_{\rm cm}-\textbf{r}/2}\big]\ket{\text{GS};\bm{\kappa}},
    \label{eq:IntermediateShiftIntrinsic}
\end{align}
which, as we will see below, can be delineated into three physically meaningful terms. In obtaining Eq.~(\ref{eq:IntermediateShiftIntrinsic}) we have recalled the transformation properties from the previous section. 

\textbf{Wannier center part:} The first term is obtained by acting $\nabla_{\bm{\kappa}}$ on $\psi^{\bm{\kappa}}_{\bm{Q}}(\textbf{r})$, which yields:
\begin{align}
\sum\limits_{\textbf{r}}\textbf{r}|\psi_{\bm{Q}}(\textbf{r})|^2\equiv \sum\limits_{\textbf{r}}\rho(\textbf{r},\textbf{r})\textbf{r},
\end{align}
where we have used the exciton envelope function’s reduced density matrix $\rho(\textbf{r},\textbf{r}') = \psi^*_{\bm{Q}}(\textbf{r}) \psi_{\bm{Q}}(\textbf{r}')$. This term originates from the polarization contributions of electrons and holes localized at their respective Wannier centers and appears as a diagonal component in $\rho$. 

\textbf{Electron cloud correction:} The second term is obtained from taking derivative of $\bm{\kappa}$ with $c^{\dagger}_{\bm{\kappa};c,\textbf{R}_{cm}+\textbf{r}/2}$. After a straightforward evaluation we arrive at:
\begin{align}
&\sum\limits_{\textbf{r},\textbf{r}'}\rho(\textbf{r},\textbf{r}')e^{i\bm{Q}\cdot(\textbf{r}'-\textbf{r})/2}\int d\textbf{x}w^*_{c,\textbf{r}}(\textbf{x})(\textbf{x}-\textbf{r}')w_{c,\textbf{r}'}(\textbf{x})=\sum\limits_{\textbf{r},\textbf{r}'}\rho(\textbf{r},\textbf{r}')e^{i\bm{Q}\cdot(\textbf{r}'-\textbf{r})/2}
\bm{d}_c(\textbf{r}'-\textbf{r}),
\end{align}
where in the last equality we have noted the translational symmetry of  Wannier functions as well as 
utilized the dipole matrix element in the Wannier basis as defined in Ref.\cite{marzari1997maximally}: $\int w^*_{c,\textbf{0}}(\textbf{x})\hat{\textbf{x}}w_{c,\textbf{R}}(\textbf{x})=\bm{d}_c(\textbf{R})$.
The off-diagonal component of density matrix $\rho(\textbf{r},\textbf{r}')$ indicates the probability of electron and hole Wannier centers are $\textbf{r}$ apart and the electrons are found at $\textbf{r}'$. The $e^{i\bm{Q}\cdot(\textbf{r}'-\textbf{r})/2}$ factor is due to an effective shift of the center-of-mass coordinate due to the electron cloud. 

The physical meaning is clear: this term captures the polarization correction from the spatial spread of conduction-band Wannier orbitals with the dipole operator reflecting charge distribution. This can be confirmed by noting that if Wannier functions $w_{c,\textbf{R}}(\textbf{x})$ are strongly localized at $\textbf{R}$, this term will be vanishingly small. 

\textbf{Hole cloud correction:} The third term is obtained from taking the derivative of $\bm{\kappa}$ with $c_{\bm{\kappa};v,\textbf{R}_{cm}-\textbf{r}/2}$ producing: 
\begin{align}
&-\sum\limits_{\textbf{r},\textbf{r}'}\rho(\textbf{r},\textbf{r}')e^{i\bm{Q}\cdot(\textbf{r}-\textbf{r}')/2}\int d\textbf{x}w^*_{v,\textbf{r}}(\textbf{x})(\textbf{x}-\textbf{r})w_{v,\textbf{r}'}(\textbf{x})=-\sum\limits_{\textbf{r},\textbf{r}'} \rho(\textbf{r},\textbf{r}')e^{i\bm{Q}\cdot(\textbf{r}-\textbf{r}')/2}
\bm{d}_{v}(\textbf{r}'-\textbf{r}) .
\end{align}

Similar to the electron case, this term is the correction to polarization due to the spatial spreading of hole Wannier orbitals from the valence band. Similar to above, the $e^{i\bm{Q}\cdot(\textbf{r}-\textbf{r}')/2}$ factor arises from a an effective shift of center-of-mass coordinates.

Combining these three terms together, we arrive at the following compact expression for the exciton shift vector:
\begin{align}
\bm{\mathcal{R}}_{0\rightarrow ex}\doteq\sum\limits_{\textbf{r},\textbf{r}'}\rho(\textbf{r},\textbf{r}')\big[ \textbf{r}\delta_{\textbf{r},\textbf{r}'}+e^{i\bm{Q}\cdot(\textbf{r}'-\textbf{r})/2}
\bm{d}_c(\textbf{r}'-\textbf{r}) -e^{i\bm{Q}\cdot(\textbf{r}-\textbf{r}')/2}
\bm{d}_v(\textbf{r}'-\textbf{r})\big].
\label{eq:ShiftIntrinsicSupp}
\end{align}

Notice that in a periodic system, the relative position $\textbf{r}$ is inherently ambiguous due to the periodic boundary conditions. 
However, due to the localized nature of the exciton envelope function $\psi(\textbf{r})$ (and thus the reduced density matrix $\rho$), $\textbf{r}$ only needs to be defined near the origin: close to the boundaries the amplitude of the exciton envelope is small rendering any ambiguity arising from the position of $\textbf{r}$ exponentially suppressed. We emphasize that the expression above in Eq.~(\ref{eq:ShiftIntrinsicSupp}) only works for localized particle-hole excitations: it immediately fails for delocalized particle-hole excitations. This is directly connected to the ill-defined nature of the position operator under periodic boundary conditions~\cite{resta1998quantum}.


\section{Symmetry properties of shift vector for localized states}

In this section, we prove that the shift vector between two localized states (e.g., excitonic transitions) transforms as a vector under the spatial symmetries of a system. To this end, we first rewrite the shift vector in terms of a manifestly gauge-invariant Wilson loop operator\cite{shi2021geometric}:
$\bm{\mathcal{R}}_{0\rightarrow \text{ex}}(\bm{Q})\equiv \nabla_{\bm{\kappa}}\text{arg}\mathcal{W}^{\bm{Q}}(\bm{\kappa})|_{\bm{\kappa}=\bm{0}},$
where the Wilson loop $\mathcal{W}^{\bm{Q}}(\bm{\kappa})$ is:
\begin{align}
\mathcal{W}^{\bm{Q}}_{0\rightarrow \text{ex}}(\bm{\kappa})=\braket{\Phi_0(\bm{0})|\Phi_0(\bm{\kappa})}\bra{\Phi_0(\bm{\kappa})}\hat{V}\ket{\Phi_{\text{ex},\bm{Q}}(\bm{\kappa})}\braket{\Phi_{\text{ex},\bm{Q}}(\bm{\kappa})|\Phi_{\text{ex}}(\bm{0})}
\end{align}
This is a direct consequence of the following fact: $\braket{\Phi_0(\bm{0})|\Phi_0(\bm{\kappa})}=1+\bm{\kappa}\cdot \bra{\Phi_0}\nabla_{\bm{\kappa}}\ket{\Phi_0}+\mathcal{O}(\kappa^2)$, which leads to $\text{lim}_{\bm{\kappa}\rightarrow 0}\nabla_{\bm{\kappa}} \text{arg}(\braket{\Phi_0(\bm{0})|\Phi_0({\bm{\kappa}})})=-i\bra{\Phi_0}\nabla_{\bm{\kappa}}\ket{\Phi_0}$ (and similarly for $\Phi_{\text{ex}}$). Here we have made explicit the $\bm{Q}$ dependence of the exciton as it will also change under spatial symmetry.

Let's assume a system has $G$ as its point group symmetry, which may contain rotation $C_n$, mirror reflections $M$, other operations and 
combinations. Under a general point group symmetry operation $g\in G$, the symmetry transformation rule is:
\begin{align}
g\ket{\Phi_{\text{ex},\bm{Q}}(\bm{\kappa})}=e^{i\theta_{\text{ex}}(\bm{\kappa})}\ket{\Phi_{\text{ex},\hat{U}_g\bm{Q}}(\hat{U}_g\bm{\kappa})},
\end{align}
where we have included a phase factor $e^{i\theta_{\text{ex}}(\bm{\kappa})}$ to account for symmetry quantum number.

In order to make connection with the Wilson loop at symmetry-transformed $\bm{\kappa}'=\hat{U}_g\bm{\kappa}$, we apply the above relation to the states:
\begin{equation}
    \begin{split}
&\mathcal{W}^{\bm{Q}}_{0\rightarrow\text{ex}}(\bm{\kappa})=\braket{\Phi_0(\bm{0})|\Phi_0(\bm{\kappa})}\bra{\Phi_0(\bm{\kappa})}\hat{V}\ket{\Phi_{\text{ex},\bm{Q}}(\bm{\kappa})}\braket{\Phi_{\text{ex},\bm{Q}}(\bm{\kappa})|\Phi_{\text{ex},\bm{Q}}(\bm{0})}. \\
&=e^{-i\theta_0(\bm{0})+i\theta_{\text{ex}}(\bm{0})}\braket{\Phi_0(\bm{0})|\Phi_0(\bm{\kappa}')}\bra{\Phi_0(\bm{\kappa}')}g\hat{V}g^{-1}\ket{\Phi_{\text{ex},\hat{U}_g\bm{Q}}(\bm{\kappa}')}\braket{\Phi_{\text{ex},\hat{U}_g\bm{Q}}(\bm{\kappa}')|\Phi_{\text{ex},\hat{U}_g\bm{Q}}(\bm{0})}\\
&=e^{-i\theta_0(\bm{0})+i\theta_{\text{ex}}(\bm{0})}\overline{\mathcal{W}}_{0\rightarrow \text{ex}}^{\hat{U}_g\bm{Q}}(\bm{\kappa}'),       
    \end{split}
\end{equation}
where we have used $\bra{\Phi_0(\bm{0})}g^{-1}g\ket{\Phi_0(\bm{\kappa})}=e^{i\theta_0(\bm{\kappa})-i\theta_0(\bm{0})}\braket{\Phi_0(\bm{0})|\Phi_0(\hat{U}_g\bm{\kappa})}$ and similarly for $\ket{\Phi_{\text{ex},\bm{Q}}(\bm{\kappa})}$, and the Wilson loop $\overline{\mathcal{W}}(\bm{\kappa}')$ is defined with the symmetry-transformed potential $g\hat{V}g^{-1}$. Notice that $\bar{\mathcal{W}}(\bm{\kappa}')$ is identical to $\mathcal{W}(\bm{\kappa}')$ apart from a symmetry-transformed $g\hat{V}g^{-1}$, but the $\hat{V}$-insensitivity of the shift vector is precisely what we have proved in the case of localized states. More precisely, $\text{arg}(\bra{\Phi_0(\bm{\kappa}')}\hat{V}_1\ket{\Phi_n(\bm{\kappa}')}\bra{\Phi_n(\bm{\kappa}')}V_2\ket{\Phi_0(\bm{\kappa}')})$ is $\bm{\kappa}'$-independent for $\hat{V}_1=g\hat{V}g^{-1}$ and $\hat{V}_2=\hat{V}$. Note that this fact {\it still holds} for excitons formed from a Chern band when we represent the states in terms of hybrid Wannier functions localized along the direction of $\bm{\kappa}'$-a point we will discuss in the following section. With these in mind, we have:
\begin{align}
    \nabla_{\bm{\kappa}'}\text{arg}(\overline{\mathcal{W}}^{\hat{U}_g\bm{Q}}_{0\rightarrow \text{ex}}(\bm{\kappa}'))\doteq \nabla_{\bm{\kappa}'}\text{arg}(\mathcal{W}^{\hat{U}_g\bm{Q}}_{0\rightarrow \text{ex}}(\bm{\kappa}'))
\end{align}

Therefore the shift vector $\bm{\mathcal{R}}_{0\rightarrow \text{ex}}$ can be evaluated with a transformed $\bm{\kappa}'=\hat{U}_g\bm{\kappa}$ (noticing that that symmetry quantum numbers $e^{i\theta_{0/\text{ex}}(\textbf{0})}$ are $\bm{\kappa}$-independent):

\begin{align}
    \bm{\mathcal{R}}_{0\rightarrow\text{ex}}(\bm{Q})=\nabla_{\bm{\kappa}}\text{arg}\mathcal{W}^{\bm{Q}}(\bm{\kappa})|_{\bm{\kappa}=0}\doteq\nabla_{\bm{\kappa}}\text{arg}\mathcal{W}^{\hat{U}_g\bm{Q}}(\bm{\kappa}')|_{\bm{\kappa}=0}=\frac{\partial \bm{\kappa}'}{\partial\bm{\kappa}}\nabla_{\bm{\kappa}'}\text{arg}\mathcal{W}^{\hat{U}_g\bm{Q}}(\bm{\kappa}')|_{\bm{\kappa}'=0}=[\hat{U}_g]^{-1}\bm{\mathcal{R}}_{0\rightarrow \text{ex}}(\hat{U}_g\bm{Q}),
\end{align}
(the last equality is due to $[\frac{\partial {\kappa}'_b}{\partial{\kappa}_a}]=[\hat{U}_g]_{ba}=[\hat{U}_g^{-1}]_{ab}$ because $\hat{U}_g$ is an orthogonal matrix). Moving the $\hat{U}_g$ to the left, we obtain Eq.~\eqref{eq:shift_vector_pg_transformation} in the main-text:
\begin{align}
\hat{U}_g[\bm{\mathcal{R}}_{0\rightarrow \text{ex}}(\bm{Q})]\doteq\bm{\mathcal{R}}_{0\rightarrow \text{ex}}(\hat{U}_g\bm{Q})
\end{align}
It then naturally follows that any exciton transition shift vector $\bm{\mathcal{R}}_{0\rightarrow \text{ex}}$ that is not invariant under $g$ should be a null vector, the simplest example being the vanishing of in-plane exciton transition shift vector under $C_{3z}$ rotational symmetry. Note that this is not true for $\bm{\mathcal{R}}_{0\rightarrow \text{free}}$.

\section{Time-dependent perturbation calculation of the shift current in a many-body setting}\label{appendix:time-dependent}
We note that the many-body shift current expression in terms of a many-body shift vector was first given in Ref.~\onlinecite{resta2024geometrical}. Here in this section, we provide an alternative derivation based on time-dependent perturbation theory in the velocity-gauge. The full many-body Hamiltonian with a time-dependent vector potential $\bm{A}$ can be written as 
\begin{equation}
   \hat{H}(\bm{A})=\frac{1}{2m}\sum\limits_{i=1}^N(\textbf{p}_i-e \bm{A})^2+\hat{V}_{\rm lattice}+\hat{H}_{\rm int},
\end{equation}
where $\hat{V}_{\rm lattice}$ is the lattice potential and $H_{\rm int}$ denotes electron-electron interactions. The full Hamiltonian can be expanded as $\hat{H}(\bm{A})=\hat{H}(0)+\hat{\bm{j}}\cdot \bm{A}$ and we denote the time-dependent interaction formally as $\hat{\bm{j}}\cdot \bm{A}=\hat{V}(t)$. The only thing we need in the following derivation is that $\hat{V}(t)=\hat{V}^{\omega}e^{-i\omega t}+\hat{V}^{-\omega}e^{i\omega t}$ and $\hat{V}^{-\omega}=(\hat{V}^{\omega})^*$. Since our focus is solely on the off-diagonal correction to the density matrix, the diamagnetic contribution to the current can be safely neglected. Below we denote the eigen-states of $\hat{H}(0)$ as $\ket{n}$ ($n=0,1,\cdots$), where $n=0$ is the ground state.

The density matrix evolves under the quantum Liouville equation in the interaction picture:
\begin{equation}\label{eq:Liouville}
    \frac{d\rho_I}{dt}=\frac{i}{\hbar}[\rho_I,\hat{V}_I(t)],
\end{equation}
where operators in the interaction picture (with subscript $I$) and the Schrodinger picture (without subscript)  are related via the relation $\hat{O}_I(t)=e^{i\hat{H}(0)/\hbar}\hat{O}(t)e^{-i\hat{H}(0)/\hbar}$. Eq.\eqref{eq:Liouville} can be solved up to second order as:
\begin{equation}
    \rho_I(t)=\rho_I^{(0)}+\frac{1}{i\hbar}\int^t_{-\infty}[\hat{V}_I(t_1),\rho_I^{(0)}]dt_1+(\frac{1}{i\hbar})^2\int^t_{-\infty}\int^{t_1}_{-\infty}[\hat{V}_I(t_1),[\hat{V}_I(t_2),\rho_I^{(0)}]]dt_2dt_1,
\end{equation}
where $\rho^{(0)}$ is the unperturbed equilibrium density matrix and in the special case of zero-temperature is simply $\rho^{(0)}=\ket{0}\bra{0}$ (generalization to finite-temperature is straightforward).

The current is therefore 
\begin{equation}
\begin{split}
    &\bm{j}(t)\equiv\text{Tr}(\rho_I(t)\hat{\bm{j}}_I(t))=\text{Tr}(\rho_I^{(0)}\hat{\bm{j}}_I(t))+\frac{1}{i\hbar}\int^{t}_{-\infty}\text{Tr}(\rho_I^{(0)}[\hat{\bm{j}}_I(t),\hat{V}_I(t_1)])\\
    &+(\frac{1}{i\hbar})^2\int^t_{-\infty}\int^{t_1}_{-\infty}\text{Tr}(\rho_I^{(0)}[[\hat{\bm{j}}_I(t),\hat{V}_I(t_1)],\hat{V}_I(t_2)])dt_2dt_1+\mathcal{O}(A^3).
\end{split}
\end{equation}

The shift current is a second-order (rectified) DC current that originates from off-diagonal elements of $\rho$ and is extracted by inserting the identity $\mathbbm{1} = \sum_n \ket{n}\bra{n}$, performing the time integral, and isolating the resonant part via a small imaginary shift in the energy denominator. Focusing on the resonant (absorptive) part produces the shift current: 
\begin{equation}\label{eq:shift}
\begin{split}
&\bm{j}_{\text{shift}}=i\pi\sum\limits_{n\neq 0}\delta(E_{n0}-\hbar\omega)\big[\sum\limits_{m\neq 0}\frac{\bra{0}\hat{\bm{j}}\ket{m}}{E_{m0}}\bra{m}\hat{V}^{-\omega}\ket{n}\bra{n}\hat{V}^{\omega}\ket{0}+\sum\limits_{m\neq n}\bra{0}\hat{V}^{-\omega}\ket{m}\frac{\bra{m}\hat{\bm{j}}\ket{n}}{E_{nm}}\bra{n}\hat{V}^{\omega}\ket{0}\big]+h.c.
    \end{split}
\end{equation}

While Eq.~(\ref{eq:shift}) describes the full shift current, it is useful to rewrite it in a form that explicitly contains the shift vector. To see this we next prove the following useful many-body sum rule:
\begin{equation}\label{eq:sum_rule}
\begin{split}
&\left(\sum_{m\neq 0}\frac{\bra{0}\hat{\bm{j}}\ket{m}}{E_{m0}}\bra{m}\hat{V}\ket{n}+\sum\limits_{m\neq n}\bra{0}\hat{V}\ket{m}\frac{\bra{m}\hat{\bm{j}}\ket{n}}{E_{nm}}\right)=-\frac{e}{\hbar}\bra{0}\hat{V}\ket{n}\cdot \left[\bra{n}\nabla_{\bm{\kappa}}\ket{n}-\bra{0}\nabla_{\bm{\kappa}}\ket{0}-\nabla_{\bm{\kappa}}\text{ln}(\bra{0}\hat{V}\ket{n})\right].
\end{split}
\end{equation}

This can be proved by first noticing the identity $\bra{m}\hat{\bm{j}}\ket{n}/E_{nm}=-\frac{e}{\hbar}\bra{m}\nabla_{\bm{\kappa}}\ket{n}$ (obtained from taking derivative with $\bm{\kappa}$ of the expression $\bra{m}H\ket{n}$ and identifying $\hbar\bm{\kappa}$ with $-e\bm{A}$). Then we sum over intermediate states $\ket{m}$ via the identity $\sum\limits_{m}\ket{m}\bra{m}=\mathbbm{1}$. Finally utilizing the terms $(\nabla_{\bm{\kappa}}\bra{0})\hat{V}\ket{n}+\bra{0}\hat{V}(\nabla_{\bm{\kappa}}\ket{n})=\nabla_{\bm{\kappa}}(\bra{0}\hat{V}\ket{n})$, we arrive at the sum rule Eq.~\eqref{eq:sum_rule}. Note that the contribution $\bra{0}(\nabla_{\bm{\kappa}}\hat{V})\ket{n}$ vanishes for $n\neq 0$. 

From this we can reexpress Eq.\eqref{eq:shift} as (taking into consideration the hermitian conjugation part):
\begin{align}
    \bm{j}_{\rm shift}=-\frac{2\pi e}{\hbar}\sum\limits_n|\bra{n}\hat{V}^{\omega}\ket{0}|^2\bm{\mathcal{R}}_{0n}\delta(E_{n}-E_0-\hbar\omega)
\end{align}
where $\bm{\mathcal{R}}_{0n}\equiv i\bra{n}\nabla_{\bm{\kappa}}\ket{n}-i\bra{0}\nabla_{\bm{\kappa}}\ket{0}+\nabla_{\bm{\kappa}}\text{arg }(\bra{0}\hat{O}\ket{n})$.

\section{Discussion of topological obstructed bands}
For topologically nontrivial bands, a global smooth gauge for Bloch wavefunctions—and hence maximally localized Wannier functions—is obstructed. However, as we now explain, the previous discussions remain valid if we now use hybrid Wannier functions. We proceed by considering a $d$-dimensional system and, for convenience, insert flux along the $x$-direction. We denote the remaining momenta collectively as $\textbf{k}_{\perp}$. We then construct hybrid Wannier functions localized along $x$~\cite{sgiarovello2001electron,taherinejad2014wannier,gresch2017z2pack}:
\begin{align}
&w_{R_x,\textbf{k}_{\perp}}(\textbf{x})=\frac{1}{\sqrt{N_x}}\sum\limits_{k_x}e^{i(\textbf{k}\cdot\textbf{x}-k_xR_x)}e^{i\phi_{\textbf{k}}}u_{\textbf{k}}(\textbf{x}),
\end{align}
where the gauge freedom $e^{i\phi_{\textbf{k}}}$ is fixed to ensure localization along $x$ and $N_x$ is the number of unit-cells along $x$ direction.

Hybrid Wannier functions have well-defined charge center along $x$ direction as $\bar{x}(\textbf{k}_{\perp})\equiv \bra{w_{R_x,\textbf{k}_{\perp}}}\hat{x}\ket{w_{R_x,\textbf{k}_{\perp}}}$ due to their localization properties along $x$. The key point of topological obstruction is that the Wannier charge center can shift when $\textbf{k}_{\perp}$ is varied, resulting in an obstruction to exponentially localized Wannier function along all spatial directions. For example, for a Chern insulator with Chern number $C$, the Wannier center will shift a net amount of $Ca$ along the $x$ direction when $k_y$ changes by $2\pi$. The point, however, is that this shift is still on the order of lattice constants, and therefore is microscopically small relative to the linear dimension $L$.

A general particle-hole state in this basis can be described as:
\begin{equation}
\begin{split}
&\ket{\psi_{\bm{Q}}}_{\text{p}-\text{h}}=\frac{1}{\sqrt{N_x}}\sum\limits_{R_x,r_x,\textbf{p}_{\perp}} \psi_{\bm{Q}}(r_x,\textbf{p}_{\perp})e^{iQ_xR_x} c^{\dagger}_{c,R_x+r_x/2,\textbf{p}_{\perp}+\bm{Q}_{\perp}/2}c_{v,R_x-r_x/2,\textbf{p}_{\perp}-\bm{Q}_{\perp}/2}\ket{\text{GS}},
\end{split}
\end{equation}
where $c^{\dagger}_{c,R_x,\textbf{k}_{\perp}}\ket{0}=\int d\textbf{x} w_{c,R_x,\textbf{k}_{\perp}}(\textbf{x})a^{\dagger}_{\textbf{x}}\ket{0}.$ Projecting the Hamiltonian $\hat{H}$ onto the particle-hole basis, we obtain a BSE Hamiltonian of the form $\mathcal{H}(r_x,r_x',\textbf{p}_{\perp},\textbf{p}_{\perp}')$. The evaluation is straightforward but lengthy, therefore we only show some representative matrix elements to illustrate the idea. The kinetic term from conduction band is $e^{iQ_x(r_x'-r_x)/2}\bra{c,r_x,\textbf{p}_{\perp}+\bm{Q}_{\perp}/2}\hat{H}_{\textbf{K}}\ket{c,r_x',\textbf{p}_{\perp}+\bm{Q}_{\perp}/2}$. And the direct interaction is:
\begin{equation}
\begin{split}&\mathcal{V}_D(r_x,r_x',\textbf{p}_{\perp,1},\textbf{p}_{\perp,2})=-\sum\limits_{\Delta R_{x}} e^{iQ_x(\Delta R_{x})}\int d\textbf{x}_1d\textbf{x}_2 w_{c,\Delta R_{x}+r_{x_2}/2,\textbf{p}_{\perp,2}+\bm{Q}_{\perp}/2}(\textbf{x}_1)w^*_{c,r_{x_1}/2,\textbf{p}_{\perp,1}+\bm{Q}_{\perp}/2}(\textbf{x}_1)\\
  &\times V(|\textbf{x}_1-\textbf{x}_2|)w_{v,-r_{x_1}/2,\textbf{p}_{\perp,1}-\bm{Q}_{\perp}/2}(\textbf{x}_2)w^*_{v,\Delta R_{x}-r_{x_2}/2,\textbf{p}_{\perp,2}-\bm{Q}_{\perp}/2}(\textbf{x}_2).
\end{split}
\end{equation}

Using the property that $|w_{n,R_x,\textbf{p}_{\perp}}(\textbf{x})|< \mathcal{F}(C,a,\xi_W)e^{-|R_x-x|/\xi_W}$ for large separation $|R_x-x|$, we can proceed exactly as before and prove that the off-diagonal elements of the BSE Hamiltonian decays as $e^{-|r_x-r_x'|/\xi_W}$, with $\xi_W$ the wave-function extent along the $x$ direction. The constant factor $\mathcal{F}(C,a,\xi_W)$ accounts for possible shifts of Wannier center when $\textbf{p}_{\perp}$ is varied, but does not change the fact that all terms of the BSE become exponentially suppressed for 
large $|r_x-r_x'|$. 

Next we show that the exciton envelope function decays exponentially at large $r_x$. The asympotic behavior can be argued as follows. At very large $r_x$, we can ignore the electron-hole interaction and the BSE Hamiltonian reduces to $\epsilon_c(\hat{p}_x,\textbf{p}_{\perp})-\epsilon_v(\hat{p}_x,\textbf{p}_{\perp})$ in the hybrid coordinate $(r_x,\textbf{p}_{\perp})$, which results from the Fourier transformation of $\epsilon_c(\textbf{p})-\epsilon_v(\textbf{p})$ along the $x$ direction and can be solved by the ansatz $e^{ip_xr_x}$. Notice that $\epsilon_c(p_x,\textbf{p}_{\perp})-\epsilon_v(p_x,\textbf{p}_{\perp})=E$ can only be satisfied when $p_x$ has imaginary parts because $E$ lies within the gap. Consequently the envelope function at large $r_x$ behaves as $ e^{i\text{Re}[p_x]r_x}e^{-\text{Im}[p_x]r_x}$ is exponentially decaying. 

As a result, when flux is inserted along the $x$ direction, the Bloch function transforms as $e^{i\textbf{k}\cdot\textbf{r}}u_{\textbf{k}+\kappa_x\textbf{e}_x}(\textbf{x})$, leading to the transformation rule of the hybrid Wannier functions: $w_{R_x,\textbf{k}_{\perp}}(\textbf{x})\rightarrow e^{-i\kappa_x(x-R_x)}w_{R_x,\textbf{k}_{\perp}}(\textbf{x})$. Therefore for an arbitrary one-body short-range operator $\hat{o}$ expanded in terms of hybrid Wannier functions $\hat{o}=\sum\limits_{R_x,R_x',\textbf{p}_{\perp},\textbf{p}_{\perp}'}o_{R_x,\textbf{p}_{\perp};R_x',\textbf{p}_{\perp}'}c^{\dagger}_{R_x,\textbf{p}_{\perp}}c_{R_x',\textbf{p}_{\perp}'}$, the matrix elements transform as $o_{R_x,\textbf{p}_{\perp};R_x',\textbf{p}_{\perp}'}\rightarrow o_{R_x,\textbf{p}_{\perp};R_x',\textbf{p}_{\perp}'}e^{-i{\kappa}_x\cdot (R_x-R_x')}$ upon flux insertion. Similarly, the matrix elements of the BSE Hamiltonian transforms as: $\mathcal{H}(r_x,r_x',\textbf{p}_{\perp},\textbf{p}_{\perp}')\rightarrow e^{-i\kappa_x(r_x-r_x')}\mathcal{H}(r_x,r_x',\textbf{p}_{\perp},\textbf{p}_{\perp}')$.

In conclusion, all the essential ingredients for our proof still hold, namely:
\begin{enumerate}
    \item The BSE Hamiltonian $\mathcal{H}(r_x,r_x',\textbf{p}_{\perp},\textbf{p}_{\perp}')$ is almost diagonal in $r_x,r_x'$, meaning that the off-diagonal matrix elements decay as $e^{-|r_x-r_x'|/\xi_W}$.
    \item The exciton solution $\psi_{\bm{Q}}(r_x,\textbf{p}_{\perp})$ to the BSE is exponentially localized in $r_x$.
    \item Coupling to $\kappa_x$ is achieved as $\mathcal{H}(r_x,r_x',\textbf{p}_{\perp},\textbf{p}_{\perp}')\rightarrow e^{-i\kappa_x(r_x-r_x')}\mathcal{H}(r_x,r_x',\textbf{p}_{\perp},\textbf{p}_{\perp}')$.
\end{enumerate}

With the above preparation, we can now proceed to put the BSE Hamiltonian on a periodic lattice with linear size $L$, adopt the ansatz $\psi_{\bm{Q}}(r_x,\textbf{p}_{\perp})\rightarrow e^{-i\kappa_xr_x}\psi_{\bm{Q}}(r_x,\textbf{p}_{\perp})$ and arrive at the conclusion that the ansatz is accurate with an exponentially suppressed deviation $(N/N_x)e^{-L_x/(2\xi_{M})}$, where $N/N_x$ comes from summation over $\textbf{p}_{\perp}$. In the thermodynamic limit, the exponential factor always dominates over $N/N_x$.

Simlar to that described in the main text, we can then utilize the transformation rule of the exciton envelope function (along $x$) and the matrix elements of the operators $\hat{V}_1$ and $\hat{V}_2$ under flux insertion, we can proceed exactly as before and prove that $\text{arg}(\bra{\Phi_0(\bm{\kappa})}\hat{V}_1 (\bm \kappa)\ket{\Phi_{\text{ex}}(\bm{\kappa})}\bra{\Phi_{\text{ex}}(\bm{\kappa})}\hat{V}_2 (\kappa)\ket{\Phi_0(\bm{\kappa})})$ is $\bm{\kappa}$-independent for any $\hat{V}_1$ and $\hat{V}_2$.

Different basis choices yield the same eigenstate up to a global phase, allowing us to choose hybrid Wannier functions localized along any direction. This freedom lets us extend the proof of the exciton shift vector’s vector-like transformation to Chern-band excitons without modification.

\section{Details of numerical simulation}
We consider a model on a honeycomb lattice with sublattices $A$ and $B$ in a unit-cell. The real space periodicity vectors are chosen as $\textbf{a}_1=(1,0)$, $\textbf{a}_2=(\frac{1}{2},\frac{\sqrt{3}}{2})$ and $\textbf{d}_1\equiv \frac{\textbf{a}_1-2\textbf{a}_2}{3}$ is the spatial displacement between sites $A$ and $B$ in a unit-cell. The reciprocal vectors are chosen as $\textbf{b}_1=2\pi (1,-\frac{1}{\sqrt{3}})$, $\textbf{b}_2=2\pi (0,\frac{2}{\sqrt{3}})$. , 

The tight-binding Hamiltonian is given by:
\begin{align}
    H=\sum\limits_{j}(-t_1c_{A,\textbf{R}_j}^{\dagger}c_{B,\textbf{R}_j}-t_2c_{A,\textbf{R}_j}^{\dagger}c_{B,\textbf{R}_j+\textbf{a}_2}-t_3c_{A,\textbf{R}_j}^{\dagger}c_{B,\textbf{R}_j-\textbf{a}_1+\textbf{a}_2})+h.c.+\frac{\Delta}{2}(c_{A,\textbf{R}_j}^{\dagger}c_{A,\textbf{R}_j}-c_{B,\textbf{R}_j}^{\dagger}c_{B,\textbf{R}_j}),
\end{align}
where $\textbf{R}_j=n_1\textbf{a}_1+n_2\textbf{a}_2$ runs through all the unit-cells. Fourier transformation is defined as 
$c_{\textbf{k}}^{A(B)}=\frac{1}{\sqrt{N}}\sum\limits_{j}e^{-i\textbf{k}\cdot \textbf{R}_j}c_{A(B),\textbf{R}_j}$. After Fourier transformation, we have:
\begin{align}
H(\textbf{k})=h_x(\textbf{k})\sigma_x+h_y(\textbf{k})\sigma_y+h_z(\textbf{k})\sigma_z,
\end{align}
where: $h_x(\textbf{k})=-t_1-t_2\text{cos}(\alpha)-t_3\text{cos}(\beta)$, $h_y(\textbf{k})=t_2\text{sin}(\alpha)+t_3\text{sin}(\beta)$ and $h_z(\textbf{k})=\frac{\Delta}{2}$ and $\alpha=\textbf{k}\cdot\textbf{a}_2$, $\beta=\textbf{k}\cdot (\textbf{a}_2-\textbf{a}_1)$.

The BSE in the momentum space is: 
\begin{align}
\bra{\textbf{p},\bm{Q}}\hat{H}\ket{\textbf{p}',\bm{Q}}=\delta_{\textbf{p},\textbf{p}'}\big(\epsilon_{c}(\textbf{p}+\bm{Q}/2)-\epsilon_{v}(\textbf{p}-\bm{Q}/2)\big)-(D-X)(\textbf{p},\textbf{p}',\bm{Q}), 
\end{align}
where the direct and exchange interactions are:
\begin{align}
&\mathcal{V}_D(\textbf{p},\textbf{p}',\bm{Q})=\frac{1}{N}V(\textbf{p}-\textbf{p}')\braket{u_{c,\textbf{p}+\bm{Q}/2}|u_{c,\textbf{p}'+\bm{Q}/2}}\braket{u_{v,\textbf{p}'-\bm{Q}/2}|u_{v,\textbf{p}-\bm{Q}/2}}\\
&\mathcal{V}_X(\textbf{p},\textbf{p}',\bm{Q})=\frac{1}{N}V(\bm{Q})\braket{u_{c,\textbf{p}+\bm{Q}/2}|u_{v,\textbf{p}-\bm{Q}/2}}\braket{u_{v,\textbf{p}'-\bm{Q}/2}|u_{c,\textbf{p}'+\bm{Q}/2}}
\end{align}

For the contact potential we are considering $V(\textbf{R})=V_{\text{int}}\delta_{\textbf{R},0}$, the Fourier transformed potential is $V(\textbf{q})=\sum\limits_{j}e^{i\textbf{q}\cdot\textbf{R}_j}V(\textbf{R}_j)=V_{\text{int}}$.

The non-interacting Hamiltonian can be readily diagonalized by:
\begin{align}
&U=\sqrt{\frac{h+h_z}{2h}}\begin{pmatrix}1 &-\frac{h_x-ih_y}{h+h_z}\\
    \frac{h_x+ih_y}{h+h_z}&1\end{pmatrix}, \text{ with }U^{\dagger}HU=\begin{pmatrix}h&0\\
    0&-h\end{pmatrix}
\end{align}
where $h=\sqrt{h_x^2+h_y^2+h_z^2}$ and the dependence on $\textbf{k}$ is suppressed. 

The flux-inserted Hamiltonian is given as follows. Notice that sublattice $A$ and $B$ are located at different spatial positions, the basis vector $(c_{A,\textbf{R}},c_{B,\textbf{R}})$ picks up a phase factor after flux $\bm{\kappa}$ is inserted and becomes $(e^{i\bm{\kappa}\cdot\textbf{R}}c_{A,\textbf{R}},e^{i\bm{\kappa}\cdot(\textbf{R}+\textbf{d}_1)}c_{B,\textbf{R}})$. Consequently the flux-inserted Hamiltonian is:
\begin{align}
&H^{\bm{\kappa}}(\textbf{k})=\mathcal{U}_{\bm{\kappa}}H(\textbf{k}+\bm{\kappa})\mathcal{U}_{\bm{\kappa}}^{\dagger},\text{ where }\mathcal{U}_{\bm{\kappa}}\equiv \begin{pmatrix}
      1 &0\\
      0& e^{-i\bm{\kappa}\cdot \textbf{d}_1}
  \end{pmatrix} .
\end{align}
The flux-inserted Hamiltonian can be diagonalized by $\mathcal{U}_{\bm{\kappa}}U(\textbf{k}+\bm{\kappa})$.

The current operator in the momentum space is defined as $\bm{\hat{j}}(\bm \kappa)\equiv \frac{\partial H^{\bm{\kappa}}}{\partial{\bm{\kappa}}}$. When projected onto the band basis, it yields the matrix elements:
\begin{align}
   &\begin{pmatrix}
       \bm{j}_{cc}&\bm{j}_{cv}\\
       \bm{j}_{vc}&\bm{j}_{vv}
   \end{pmatrix}\equiv [U^{\dagger}(\textbf{k}+\bm{\kappa})\mathcal{U}_{-\bm{\kappa}}](\partial_{\bm{\kappa}}H^{\bm{\kappa}}(\textbf{k}))[\mathcal{U}_{\bm{\kappa}}U(\textbf{k}+\bm{\kappa})].
\end{align}

The matrix element corresponds to optical transition between the ground state and an exciton state can therefore be evaluated as:
\begin{align}
    &\bra{\Phi_0}\bm{j}\ket{\Phi_{\text{ex}}}=\frac{1}{\sqrt{N}}\sum\limits_{\textbf{k}}\psi(\textbf{k})\bm{j}_{vc}(\textbf{k}).
\end{align}

In order to break $C_{3z}$ symmetry and arrive at a non-zero exciton shift vector, we choose the parameters $t_2=t_3=\Delta=t_0$, $t_1=0.8t_0$ and $V_{\text{int}}=2.5t_0$. Flux is inserted along $\textbf{b}_1$ direction. The Wilson loop is defined as $\mathcal{W}=\bra{\Phi_0}\hat{V}_1\ket{\Phi_n}\bra{\Phi_n}\hat{V}_2\ket{\Phi_0}$ with $\hat{V}_1 = (\sqrt{3}\hat{j}_x + \hat{j}_y)/2$ and $\hat{V}_2 = (\hat{j}_x - \sqrt{3}\hat{j}_y)/2$. 

For each system size $L$, the standard deviation of $\text{arg}(\mathcal{W}_{\bm{\kappa}})$ is evaluated over $\kappa L/2\pi$ values ranging from $0.01$ to $1.01$ with step size $0.1$. The offset $0.01$ is added to avoid possible degeneracies at $\kappa L=0,\pi$ arising from Kramers degeneracy\cite{fu2006time}. The data in Fig.~\ref{fig:numerical_result}(a,b) are simulated on a lattice with $\sqrt{N}=31$. In producing the scaling with $\sqrt{N}$ in Fig.~\ref{fig:numerical_result}(a,c), we have taken $\sqrt{N}$ ranging from $7$ to $49$ with steps of $6$.

\end{widetext}
\end{document}